\documentclass[pageno]{jpaper}

\usepackage{times}

\usepackage{datetime}
\usepackage{url}
\usepackage[dvipdfm]{hyperref}	
\hypersetup{citecolor=blue}	
\usepackage[normalem]{ulem}
\usepackage{graphicx}
\graphicspath{{./figures/}}
\usepackage{url}
\usepackage{multirow}

\usepackage{subfig}

\usepackage{authblk}

\newcommand{\email}[1]{\small \href{mailto:#1}{\nolinkurl{#1}} } 

\begin{document}

\title{Isolate First, Then Share: a New OS Architecture for Datacenter Computing}

\author[$\star$,$\ast$]{Gang Lu}
\author[$\ast$,$\ddagger$]{Jianfeng Zhan}
\author[$\star$]{Chongkang Tan}
\author[$\star$]{Xinlong Lin}
\author[$\ast$,$\ddagger$]{Defei Kong}
\author[$\ast$]{Chen Zheng}
\author[$\ast$,$\ddagger$]{Fei Tang}
\author[$\ast$,$\ddagger$]{Cheng Huang}
\author[$\ast$]{Lei Wang}
\author[$\ast$,$\ddagger$]{Tianshu Hao}
\affil[$\star$]{Beijing Academy of Frontier Science and Technology}
\affil[$\ast$]{Institue of Computing Technolgy, Chinese Academy of Sciences}
\affil[$\ddagger$]{University of Chinese Academy of Sciences}
\affil[ ]{\email{lugang@mail.bafst.com}, \email{zhanjianfeng@ict.ac.cn}, \email{tanchongkang@ict.ac.cn}, \email{linxinlong@mail.bafst.com},  \email{kongdefei@ict.ac.cn}, \email{zhengchen@ict.ac.cn}, 
\email{tangfei@ict.ac.cn}, \email{huangcheng@ict.ac.cn}, \email{wanglei_2011@ict.ac.cn}, \email{haotianshu@ict.ac.cn}}

\date{}

\maketitle

\thispagestyle{empty}

\begin{abstract}

This paper presents the "\emph{isolate first, then share}" OS model in which the machine's processor cores,
memory, and devices are divided up between disparate OS instances, and a new abstraction---subOS---is proposed to encapsulate an OS instance that can be created, destroyed, and resized on-the-fly.  
The intuition is that this avoids shared kernel states between applications, which in turn
reduces performance loss caused by contention.  
We decompose the OS into the supervisor and several subOSes \emph{running at the same privilege level}:  
 a subOS directly manages physical resources, 
  while the supervisor can  
 create, destroy, resize a subOS on-the-fly.
The supervisor and subOSes have few state sharing, but  fast inter-subOS communication mechanisms are provided
   on demand.

We present the first implementation---RainForest, which supports unmodified Linux binaries.
Our comprehensive evaluation shows RainForest outperforms Linux with four different kernels, LXC, and Xen in terms of worst-case  and average performance most of time when running a large number of benchmarks.
 The source code is available from \emph{deleted for double-blind review}.

\end{abstract}

\section{Introduction} \label{section_introduction}

 The shift  toward datacenter computing (in short, DC), e.g., warehouse-scale computing~\cite{barroso2009datacenter, barroso2013datacenter} and cloud computing~\cite{shue2012performance, sekar2011verifiable, agmon2012resource, mogul2013nic, kooburat2011best, Madhavapeddy:2013:ULO:2451116.2451167} calls for the new OS architecture.

  On one hand, previous joint work between academia and industry~\cite{kanev2015profiling} shows modern DC workloads are with significant diversity
in workload behavior with no single silver-bullet application to optimize for and with no major intra-application hotspots.
  On the other hand,   
modern DC workloads have differentiated QoS requirements in terms of  the average  or worst-case performance.
 As a notable emerging class of workloads, user-facing services require very low tail latency~\cite{dean2013tail, kasture2014ubik, janapa2010web}---worst-case performance---instead of average performance on each node.
When a latency-critical workload is mixed with others, if the OS can not improve tail latency, the resource utilization will be low.
Meanwhile, many other workloads face the  scalability challenges in terms of average performance~\cite{unrau1995hierarchical, gough2007kernel, guniguntala2008read, mellor1991algorithms, russinovich2008inside, clements2015scalable, Baumann:2009:multikernel}.

 This observation has two implications for the OS research efforts.
 First, it is no longer  possible to tackle this challenge in  an ad-hoc manner by repeatedly identifying and removing the bottlenecks~\cite{unrau1995hierarchical, gough2007kernel, guniguntala2008read, mellor1991algorithms, russinovich2008inside}. Second, even we can optimize the OS for the main classes of workloads, if they require different kernel changes, there is no guarantee that the changes will compose~\cite{Belay:2012:DSU:2387880.2387913}.

 Traditional OS architectures are mainly proposed for the average performance~\cite{Saltzer:1974:PCI:361011.361067, Graham:1968:PIP:363095.363146, Chase:1994:SPS:195792.195795, Howry:1972:MSP:850614.850619, banga1999resource, barham2003xen}.
The widely used Linux, Linux container (LXC)~\cite{banga1999resource}, or Xen~\cite{barham2003xen}, adopt monolithic kernels or virtual machine monitor (VMM) that share numerous data structures protected by locks (\emph{share first}), and then a process or virtual machine (VM) is proposed to guarantee performance isolation (\emph{then isolate}). In the rest of this paper, we call this kind of OS architectures "\emph{share first, then isolate}" (in short, SFTI).
 In addition to its ad-hoc manner by repeatedly identifying and removing the  (average) performance bottlenecks, the SFTI OS architecture has its inherent structure obstacle in achieving the worst-case performance: contending shared structures in the kernel deteriorates performance outliers. Previous state-of-the-art systems have mainly focused on improving the scalability of the existing model in terms of the average performance (e.g., sv6~\cite{clements2015scalable}, BarrelFish~\cite{Baumann:2009:multikernel}),  or on tackling tail latency reduction challenges (e.g., IX~\cite{belay2014ix}, Arrakis~\cite{peter2014arrakis}).

 \emph{Our goal is to build a new OS architecture that not only gracefully achieves disparate performance goals in terms of both average and worst-case performance, but also protects software investment}.
 While previous work done on exokernels~\cite{peter2014arrakis, Baumann:2009:multikernel, belay2014ix, Belay:2012:DSU:2387880.2387913, boyd2008corey}
offers interesting alternative approaches to either scalability or tail latency reduction challenges, monolithic kernels are  commercially significant~\cite{banga1999resource}, especially for DC workloads.




  \begin{figure}[t]
\setlength{\abovecaptionskip}{3pt}
\setlength{\belowcaptionskip}{0pt}
  \centering
  \includegraphics[scale=0.50]{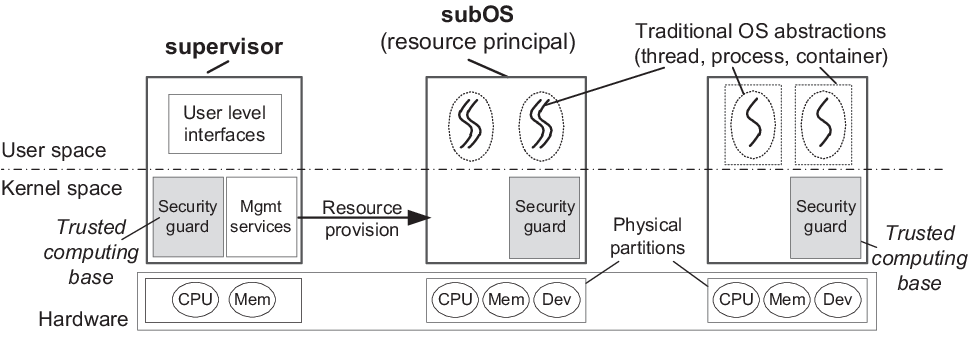}
  \caption{The IFTS OS architecture.}
  \label{horizontal_os_model}
\end{figure}

\begin{figure*}[t]
\setlength{\abovecaptionskip}{0pt}
\setlength{\belowcaptionskip}{3pt}
    \centering
    \includegraphics[width=0.98\linewidth]{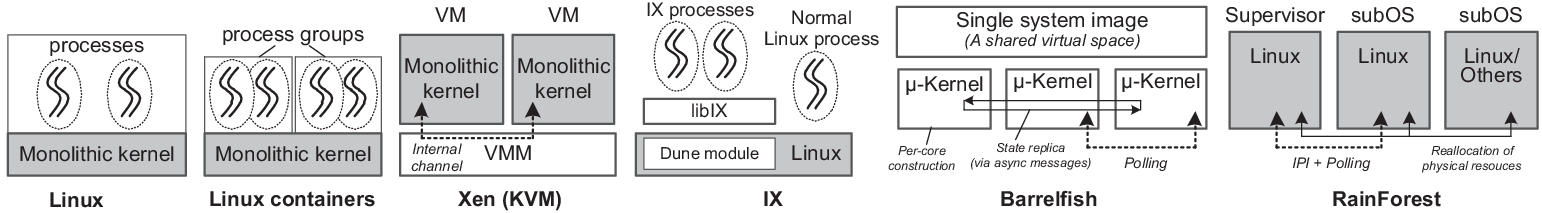}
    \caption{Structural comparison of RainForest with related systems.}
    \label{fig_related}
\end{figure*}

This paper presents an "isolate first, then share" (in short, IFTS) OS model guided by three design principles: \emph{elastic OS instance}; \emph{run at the
same privilege level}; and\emph{ confined state sharing}.
We divide up the machine's processor cores, memory, and devices among disparate OS instances,
and propose a new OS abstraction---subOS---encapsulating an OS instance that can be created, destroyed, and resized on the fly (\emph{elastic OS instance});
We decompose the OS into the supervisor and subOSes \emph{running at the same privilege level}:  
a subOS  directly  manages physical resources and runs applications, while the supervisor enables resource sharing through creating, destroying, resizing a subOS on-the-fly; 
 The supervisor and subOSes have few state sharing, but fast inter-subOS communication mechanisms based on shared memory and IPIs are provided on demand (\emph{confined state sharing}).
As the supervisor and subOSes run in the same privilege level, we take a software approach to enforce security isolation.
We propose Security Guard (in short SG)---an added kernel module in both the supervisor and each subOS kernel space utilizing the mutual protection of Instrumentation-based Privilege Restriction (IPR) and Address Space Randomization (ASR)~\cite{Deng:2017:DWT}.   IPR intercepts the specified privileged
operations in the supervisor or subOS kernels and transfers these operations to
SG, while ASR hides SG in the address
space from the supervisor or subOS kernels, which is demonstrated to be secure with trivial overhead~\cite{Deng:2017:DWT}.

On several Intel Xeon platforms, we applied the IFTS OS model to build the first working prototype---RainForest on the basis of Linux 2.6.32. Figure~\ref{horizontal_os_model} shows the IFTS OS architecture.
\emph{Our current implementation does not include safety isolation.}
We performed comprehensive evaluation of RainForest against four Linux kernels: 1) 2.6.32; 2) 3.17.4; 3) 4.9.40; 4) 2.6.35M,
a \underline{m}odified version of 2.6.35 integrated with sloppy counters~\cite{Boyd-Wickizer:2010:MOSBench}; LXC (version 0.7.5 ); XEN (version 4.0.0).  Our comprehensive evaluations show RainForest outperforms Linux with four different kernels, LXC, and Xen in terms of worst-case and average performance most of time when running a large number of benchmarks.

\section{Related Work}


\begin{figure*}[t]
\setlength{\abovecaptionskip}{0pt}
\setlength{\belowcaptionskip}{3pt}
    \centering
    \subfloat[]{\label{fig_motivation_mmap}\includegraphics[width=0.267\linewidth]{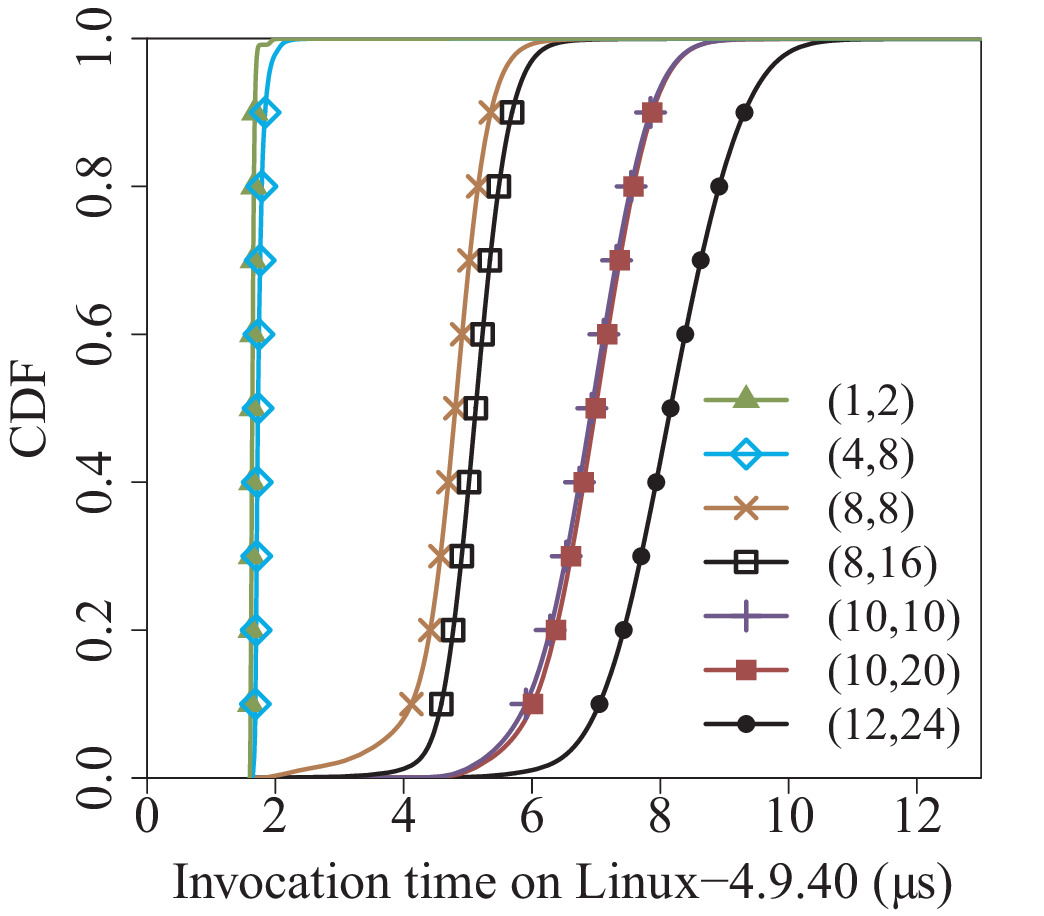}}
    \subfloat[]{\label{fig_motivation_tail_latency_search}\includegraphics[width=0.675\linewidth]{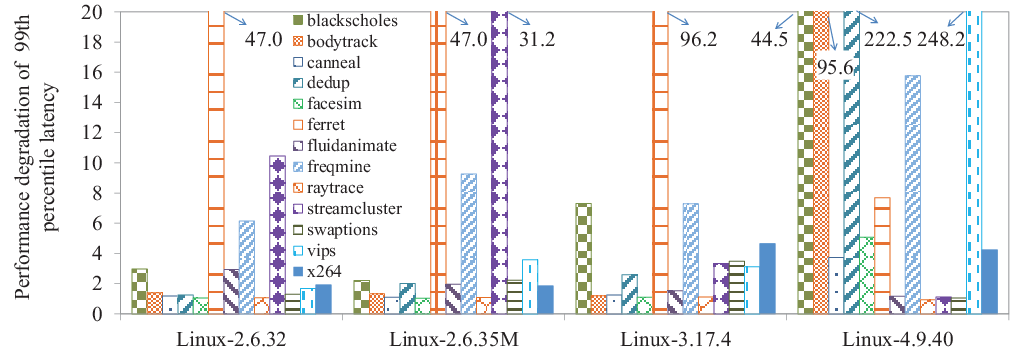}}
    \caption{\ref{fig_motivation_mmap} shows the cumulative latency distribution of \emph{mmap} from Will-it-scale~\cite{Will-It-Scale} on a 12-core server. (x, y) indicates $y$ processes running on $x$ cores.
    \ref{fig_motivation_tail_latency_search} shows the tail latency of Search is slowed down \emph{n} times (in the Y axis) when co-located with each background PARSEC benchmark. The tail latency of \emph{Search} at 300 req/s on a 12-core server with four Linux kernels  is 108.6, 134.5, 127.7 ms, 138.1 ms, respectively.}
    \label{fig_motivation_tail_latency}
\end{figure*}

 RainForest not only gracefully achieves disparate performance goals in terms of \emph{both average and worst-case performance}, but also protects software investments.
The closest system is Solaris/Sparc~\cite{SUN_Dynamic_system_domains}, which runs multiple OS instances. 
The firmware hypervisor beneath logical domains runs in an additional hyper-privileged mode, and it handles all hyper-privileged requests from logical domains via hypervisor APIs. While in RainForest, the supervisor and several subOSes run at the same privilege level, and a subOS directly manages resources without the intervention from the supervisor. Figure~\ref{fig_related} performs the structural comparison of RainForest with other related systems.


Most of previous systems focus on \emph{the average performance}.
 First, 
  much previous work scales the existing monolithic OSes to accommodate many processors in an ad-hoc manner
~\cite{unrau1995hierarchical, gough2007kernel, guniguntala2008read, mellor1991algorithms, russinovich2008inside}.  Focusing on how to reduce sharing for shared-memory systems, Tornado~\cite{gamsa1999tornado} and K42~\cite{krieger2006k42} introduce clustered objects to optimize data sharing
 through the balance of partitioning and replication. Resource containers/LXC~\cite{banga1999resource} separates a protection domain from a resource principal but shares the same structure of a monolithic kernel.

  Second, microkernels~\cite{Ford:1996:MMR:238721.238769, batlivala1992experience, golub1990unix, engler1995exokernel, Elphinstone:2013:LSW:2517349.2522720} move most kernel management into user space services, but globally-shared data structures exist in kernel components. SawMill~\cite{gefflaut2000sawmill}, Mach-US~\cite{julin1995mach}, Pebble~\cite{gabber1999pebble}, and MINIX~\cite{herder2006minix} decompose the OS into a set of user-level servers executing on the globally-shared microkernel, and leverage a set of
services to obtain and manage resources locally.
  A recent system---Tessellation~\cite{Colmenares:2013:TRO:2463209.2488827} factors OS services  and implements them in user space (similar to fos~\cite{wentzlaff2009factored}).  For the multi-server OSes~\cite{gefflaut2000sawmill, julin1995mach, gabber1999pebble, herder2006minix}, the OS is decomposed into several servers or services with complex dependencies, which amplifies the performance outliers in serving a user-facing request. In addition, multi-server OSes have a globally-shared micro-kernel,  while each subOS in RainForest  runs independently most of time.

Third, OS functionalities of an exokernel~\cite{Engler:1995:EOS:224056.224076} are pushed to library OSes linked with individual user-level processes. The globally-shared kernel still controls the access of low-level resources, handles scheduling, interrupts, and exceptions, and delivers network packets.
Corey~\cite{boyd2008corey} resembles an exokernel which defines a shared abstraction that allows applications to dynamically create lookup tables of kernel objects and determine how these tables are shared.  


Fourth, there are a number of other ways to build virtualization systems, which
   range from the hardware (e.g., IBM LPARs~\cite{Jann:2003:IBM_LPARs}) up to the full software, including hardware abstraction layer VMs (e.g., Xen~\cite{barham2003xen}), operating system layer VMs (e.g., LXC/Docker~\cite{banga1999resource, merkel2014docker}, VServer~\cite{soltesz2007container}), and hosted VMs (e.g., KVM)~\cite{kivity2007kvm}.
In Xen, even physical resources are directly allocated to a VM, it still needs VMM intervention for privileged operations and  exceptions handling. Differently, in RainForest, the machine's cores,
memory, and devices are divided up among subOSes, and the supervisor and OSes run at the same privilege level. Recursive virtual machines~\cite{Ford:1996:MMR:238721.238769, goldberg1974survey} ~\cite{zhang2011cloudvisor} allow OSes to be decomposed vertically by implementing OS functionalities in stackable VMMs.
Disco and Cellular Disco~\cite{govil1999cellular, bugnion1997disco}
use virtual machines to scale commodity multi-processor OSes.            
VirtuOS~\cite{nikolaev2013virtuos} exploits virtualization to isolate
and protect vertical slices of existing OS kernels.
Unikernels~\cite{Madhavapeddy:2013:ULO:2451116.2451167} are single-purpose appliances that are compile-time specialized into standalone kernels and sealed against modification when deployed to a cloud platform.

Finally, a number of systems reconstruct the traditional OS as a distributed system: either base on hardware partitioning~\cite{SUN_Dynamic_system_domains}, or view the OS states as replicated
~\cite{Baumann:2009:multikernel, schupbach2008embracing, beckmann2014pika, barbalacetowards, shimosawa2008logical,kale2011distributing, nomura2011mint}, or span an OS across multiple cache-coherence domains~\cite{lin2014k2} or hardware accelerators~\cite{nightingale2009helios},
or restrict the management functions of the hypervisor~\cite{keller2010nohype, szefer2011eliminating}.
 BarrelFish reconstructs the OS as a peer-to-peer distributed system built on exokernels, whose states are strictly synchronized via message passing among all CPU drivers on each core. 
Popcorn~\cite{Barbalace:2015:popcorn} constructs a single system image on top of multiple Linux kernels (each runs on a different ISA) through a compiler framework, making applications run transparently amongst different ISA processors. Differently, the IFTS OS model lets
applications running on different subOSes to avoid the cost of sharing if they do not intend to do so. In addition, the supervisor is in charge of resource management, while each subOS can be created, destroyed, and resized on the fly.

Focusing on tail latency reduction, the recent two operating systems, Arrakis~\cite{peter2014arrakis, Baumann:2009:multikernel}, and IX ~\cite{belay2014ix, Belay:2012:DSU:2387880.2387913} make applications directly access virtualized and physical I/O devices  through customized service servers and libos, respectively, so as to speed up network-intensive workloads. The subOS abstraction is orthogonal to the existing OS abstraction like DUNE/IX, and we can integrate the DUNE/IX module into subOSes specifically running latency-critical workloads, as the latter is a monolithic Linux kernel. Other tail-latency tolerant systems~\cite{kasture2014ubik, leverich2014reconciling, litales, jalaparti2013speeding, vulimiri2012more, ravindranath2013timecard, zhang2013cpi} tackle this challenges from perspective of architecture or middleware, or scheduling policy.

 Other related  work includes performance isolation solutions~\cite{verghese1998performance, shue2012performance, Gulati:2010:MHT:1924943.1924974, Ghodsi:2011:DRF:1972457.1972490}, using queuing policies or resource allocation algorithms, and resource accountability~\cite{chen2013towards, sekar2011verifiable, agmon2012resource}.

\section{Motivation} \label{Section_Motivation}

In this section, the hardware and software configurations are consistent with that in Section~\ref{section_evaluation} accordingly.

\subsection{Tail latency reduction challenge}\label{section_motivation_tail_latency}

Figure ~\ref{fig_motivation_mmap} shows on a recent Linux kernel (4.9.40) the cumulative distribution of latencies of a simple micro-benchmark---\emph{mmap} (\emph{the average is about  few microseconds}) from the Will-it-scale benchmark suite~\cite{Will-It-Scale} on a 12-core server. This benchmark creates multiple processes, each creating a 128M file and then repeatedly mapping and unmapping it into and from the memory. On the same number of cores, we have the following observation: as the process number  increases, tail latency  significantly deteriorates.
 It is mainly because of frequent accesses on a global variable \emph{vm\_commited\_as} protected by a spinlock.  vm\_commited\_as is a percpu\_counter and should streamline  concurrent updates by using the local counter in vm\_commited\_as. But once the update is beyond the \emph{percpu\_counter\_batch} limit, it will overflow into the global count, causing false sharing and high synchronization overhead.
Our experiments show similar trends in other Will-It-Scale benchmarks (e.g., pagefault, malloc). 

\subsection{Performance interferences of co-located workloads}\label{section_motivation_isolation}
Figure~\ref{fig_motivation_tail_latency_search} measures the deterioration of the tail latency of \emph{Search}---a search engine workload from the BigDataBench benchmark suite~\cite{Wang:2014:BigDataBench} (\emph{its average latency is about tens or hundreds of milliseconds}) co-located with each PARSEC workload~\cite{PARSEC_Source_code} on a 12-core server running four kernels: 2.6.32, 2.6.35M~\cite{Boyd-Wickizer:2010:MOSBench}, 3.17.4, and 4.9.40.  Although consolidation increases the resource utilization, the tail latency is slowed down tens of times, which is unacceptable for large-scale services. Even we fix  CPU affinities for Search and PARSEC processes and specify the \emph{local allocate} memory allocation policy, the tail latency is maximally slowed down by above forty times. 

\subsection{Root cause discussion}\label{motivation_discussion}

The literatures~\cite{litales, Boyd-Wickizer:2010:MOSBench, gough2007kernel} conclude the microscopic reasons for the performance outliers residing on both software and hardware levels: (1) contention for shared locks, i.e., locks for virtual memory operations, process creation and etc.; (2) competing for software resources, such as caches, lists, and etc.; (3) other in-kernel mechanisms like scheduling and timers; (4) micro-architecture events for consistency, such as TLB shootdown, cache snooping, etc.; and (5) competing limited hardware capacities, such as bandwidth.

We can contribute most of the microscopic reasons (1, 2, 4) to the SFTI architectures that share resources and states from bottom to up.
Since the computers are embracing extraordinary densities of hardware and DC workloads become increasingly  diverse, large scale and sensitive to interferences, we believe it is the right time to reconstruct the OS.



\section{The IFTS OS Model}\label{section_os_model}

The IFTS OS model is guided by three design principles.

\subsection{Elastic OS instance}


 In the IFTS OS model, the machine's cores,
memory, and devices are divided up between disparate OS instances.  
     We propose \emph{subOS} as a new OS abstraction that encapsulates an OS instance.
     Subject to the hardware architecture, each subOS may include at least a CPU core or SMT (Simultaneous Multi-Treading) thread (if supported), a memory region at a fine or coarse granularity, and optionally several peripherals.




This "isolate first" approach lets
applications avoid the cost of sharing if they do not intend to do so.
Instead, in a system with global sharing of kernel structures, processes contend with each other for limited resources, such as dentry cache, inode cache, page cache, and even routing table. 
 Note that we do not preclude subOSes sharing memory between cores (see Section~\ref{subsubsection_state_sharing}).
 On the other hand, hardware units like TLB entries are no longer necessarily shared by subOSes, and TLB shootdown is unnecessary to broadcast to all cores.

Physically partitioning node  resources among subOSes makes resource sharing much difficult.  To overcome this limitation, we allow creating, destroying and resizing a subOS on the fly, which we call \emph{elastic OS instance}. This is an important extension to reflect the requirements of DC computing. Their workloads often  change significantly in terms of both load levels and resource demands, so an IFTS OS needs to adjust the subOS number or resize each subOS's resources.
This property not only facilitates flexible resource sharing, but also supports shrinking durations of rental periods and increasingly fine grained resources offered for sale~\cite{agmon2012resource}. 

Most of server workloads are fast changing, e.g., their ratios of peak loads to normal loads are high. So resizing a subOS must be fast or else it will  result in significant QoS violation or throughput loss.
Experiences show the feasibility of hot-adding or hot-plugging a physical device in Linux~\cite{mwaikambo2004linuxCPUhotplug, schopp2006memoryhotplug, kroah2001deivcehotplug} or decoupling cores from the kernel in Barrelfish/DC~\cite{Zellweger:2014:DecouplingCores}.
We can further shorten the long path of initializing or defunctioning the physical resource from the ground up in existing approaches as discussed in Section~\ref{implementation}.



Resource overcommitment increases hardware utilization in VMM as most VMs use only a small portion of the physical memory that is allocated to them. Similarly, an IFTS OS allows resource preemption among subOSes. 
 For example, a subOS preempts cores of others if its load is much heavier. Latency critical workloads with a higher priority definitely  benefit from resource preemption when mixed with offline batches.

 SubOS is the main abstraction in our IFTS OS model.
 First, the subOS abstraction makes resource accounting much accurate. A subOS owns exclusive resources, independently runs applications, and shares few states with others, which clears up the confusion of attributing resource consumptions to applications because of  scheduling, interrupt serving, peripheral drivers and etc.

Second, the subOS abstraction is proposed as an OS management facility. 
We can flexibly  provision physical resources through creating, destroying, or resizing a subOS.  From this perspective, the subOS abstraction is orthogonal to the existing OS abstractions proposed for resource management, i.e. the resource container~\cite{banga1999resource}, and the DUNE/IX abstraction~\cite{Belay:2012:DSU:2387880.2387913, belay2014ix}.
Third, two subOSes can establish internal communication channels, which resembles inter-process communication (IPC) in a monolithic OS. A subOS can map a memory zone of another subOS to enable shared memory like \emph{mmap} in Linux. 


\subsection{Run at the same privilege level}

\begin{table*}[t]
\renewcommand{\arraystretch}{1.1}
\setlength{\abovecaptionskip}{3pt}
\setlength{\belowcaptionskip}{0pt}
\setlength{\textfloatsep} {0pt plus 2pt}
\setlength{\tabcolsep}{5pt}
\centering
\caption{Shared states compared among different operating systems.}
\label{tab_confined_states}
\newsavebox{\tablebox}
\begin{lrbox}{\tablebox}
\begin{tabular}{|c|p{200pt}|p{120pt}|p{160pt}|}
\hline
\textbf{System} & \textbf{Structures protected by locks} & \textbf{Shared software resources} & \textbf{Micro-architecture level}	\\ \hline
Linux & $dentry\ reference$, $vfsmount$, $vm\_commited\_as$, etc.~\cite{Boyd-Wickizer:2010:MOSBench}
		 & $dentry$, $inode$, $vfsmount\ table$, etc.~\cite{Boyd-Wickizer:2010:MOSBench}
		 & Events across all cores: TLB shootdown, LLC snooping, etc. \\ \hline
Linux containers &  Locks in kernel and introduced by cgroups, i.e., page\_cgroup lock~\cite{Ahn:2016:IIR}, request-queue lock~\cite{Huang:2016:ESL}
		& Same as Linux
		& Events across all cores: same as Linux; cache conflict in cgroups subsystems\\ \hline
Xen/VMM & file locks~\cite{Nitu:2017:SBQ}, many spin-locks operations~\cite{Zhong:2012:OXH}, etc.
		& Xenstore, a shared translation array, event channels, etc.~\cite{barham2003xen}
		& Events across all cores: same as Linux. \\ \hline
IX &   Locks in Linux kernel and user space driver
		& None (Static allocation)
		& Events across all cores \\ \hline
Barrelfish/Arakis & All states are globally consistent through one-phase or two-phase commit protocol~\cite{Baumann:2009:multikernel}
		& None
		& Events are limited in a single core \\ \hline
RainForest & Resource Descriptions of a subOS (lock-free)
		& None
		& Events are limited within a subOS \\ \hline
\end{tabular}
\end{lrbox}
\scalebox{0.8}{\usebox{\tablebox}}
\end{table*}

We   decompose the OS into the supervisor and several subOSes \emph{running at the same privilege level}: the supervisor discovers, monitors, and provisions resources,  while each  subOS independently manages resources and runs applications.

The roles of the supervisor and subOSes are explicitly differentiated. The supervisor gains a global view of the machine's physical resources and allocates resources for a subOS. A subOS initializes and drives the resources on its own and creates other high-level OS abstractions for better resource sharing within a subOS, e.g., process, thread, and resource container.
  A subOS becomes more efficient not only because of minimum involvement from the supervisor, but also because it  knows   resource management better than the supervisor. In addition,
the transient failure of the supervisor or the other subOSes will not lead to
the failure of the subOS, vice versa.

However,  threats come from the bugs and vulnerabilities within the kernel of the supervisor or subOSes. Since the subOS or supervisor kernel has the highest privilege, attackers who compromise the kernel could manipulate the underlying hardware state  by executing privileged operations.
 To avoid such threat, we take a software approach
and propose SG---an added kernel module in both the supervisor and each subOS kernel space---which  utilizes the mutual protection of Instrumentation-based Privilege Restriction (IPR) and Address Space Randomization (ASR)~\cite{Deng:2017:DWT} to enforce security isolation among the supervisor and subOS kernels.
ASR is a lightweight approach to randomize the code and data locations of SG in the supervisor and subOSes kernels  to ensure their locations impossible to guess in the virtual address space, thus preventing the adversary who has knowledge of the memory locations from subverting them.
Through IPR, we can intercept the specified privileged operations in the kernel and transfer them to SG in the same privilege level.




SG in each subOS kernel maps the resource descriptions of each subOS into its own address space as read-only. And only SG in the supervisor kernel has the right to modify it. SG only needs to read the descriptions to check the legality of privileged operations in the kernel. On one hand, it intercepts the operation like $mov-to-cr3$ and page table update to refrain the supervisor and subOSes from accessing invalid physical address space.
On the other hand, it intercepts the sensitive interrupt instructions, e.g., HALT and start-up IPI. Besides, intercepting the IOMMU configuration is essential if we need guarantee DMA isolation and interrupt safety.

As the code size of SG is small, we consider it as trusted computing base (TCB) while the other kernel code and user-space applications are untrusted. The integrity of TCB can be ensured using the authenticated boot provided by the platform module (TPM)~\cite{TPM}.
\subsection{Confined state sharing}\label{subsubsection_state_sharing}

In the IFTS OS model, the supervisor and subOSes have few state sharing, but  fast inter-subOS communication mechanisms are provided
   on demand.



Physically partitioning node resources inherently reduces the sharing scope  inside a subOS. Within a subOS, software operations like lock/unlock and micro-architectural operations like TLB shootdown and cache synchronization will thus happen within a smaller scope. Meanwhile, the traffic through system interconnects will also decline. However, extremely restricting a subOS onto a single core is probably not the best option. Applications may expect for a different number of cores and thread-to-core mappings~\cite{Tang:2011:IMS}.

Inter-subOS state sharing is kept on a very low level. It is ideal that no state is shared among subOSes.
But the hardware trend towards high density and fast interconnection accelerates intra-node data exchange. And the supervisor has to communicate with subOSes for coordination.
On one hand, we demand the data structures shared by the supervisor and all subOSes are quite limited. As resource management is locally  conducted by subOS kernels, global coordination and monitoring enforced by the supervisor are much reduced. On the other hand, we facilitate  two subOSes to construct mutual communication channels on demand. In this context, the negative effects of frequent updates in the communication are neutralized by  a small sharing degree. It also makes sense in the  scenario that the supervisor transfers messages to a subOS for dynamic resource reconfiguration. 
The IFTS model encourages  making a tradeoff of resource sharing  within a subOS and among subOSes, which gains the best performance for typical server workloads, e.g., the  Spark workload as shown in Section~\ref{evaluation_spark}.



Table~\ref{tab_confined_states} summarizes RainForest's state sharing reduced against other systems, including Linux, XEN, LXC, BarrelFish/Arrakis, and IX. We investigated the shared states  inside an OS from three aspects: contention for shared locks, competing software resources, and micro-architecture events for consistency. We can see that only RainForest has the least shared states and the micro-architecture events are within a subOS. As for the communication channels between two subOSes, they are established on demand and the shared structures do not have influence on other subOSes.

\section{The Implementation}\label{implementation}
We explore the implications of these principles by describing the implementation of the RainForest OS, which is not the only way to apply the IFTS OS model.

\begin{table}[t]
\renewcommand{\arraystretch}{1.1}
\setlength{\abovecaptionskip}{3pt}
\setlength{\belowcaptionskip}{0pt}
\setlength{\textfloatsep} {0pt plus 2pt}
\setlength{\tabcolsep}{5pt}
\centering
\caption{Hardware configurations of two servers.}
\label{table_server_config}
\begin{lrbox}{\tablebox}
\begin{tabular}{|c|c|c|}
\hline
Type & S-A & S-B \\ \hline
CPU & Intel Xeon E5645 & Intel Xeon E7-8870 \\ \hline
Number of cores & 6 cores@2.4GHz & 10 cores@2.4GHz \\ \hline
Number of threads & 12 & 20 \\ \hline
Sockets & 2 & 4 \\ \hline
L3 Cache & 12 MB & 30 MB \\ \hline
DRAM capacity & 32 GB, DDR3 & 1 TB, DDR3\\ \hline
NICs & \multicolumn{2}{c|}{8 Intel 82580 ethernet 1000Mb/s} \\ \hline
HDDs & \multicolumn{2}{c|}{8 Seagate 1TB 7200RPM, 64MB cache} \\ \hline
\end{tabular}
\end{lrbox}
\scalebox{0.8}{\usebox{\tablebox}}
\end{table}

\subsection{Targeted platforms}~\label{section_hardware_platform}
In this paper, we focus on applying the IFTS OS model on a cc-NUMA (cache-coherent non-uniform memory architecture) server with a single PCIe root complex. On such server, we assume there are abundant physical resources divided up among subOSes. Table~\ref{table_server_config} shows the configuration details of two different servers: a mainstream 12-core server (\emph{S-A}) and a near-future mainstream server with 40 cores and 1 TB memory (\emph{S-B}). In the rest of this paper, the performance numbers are reported on these servers.


\subsection{System structure}\label{section_system_structure}


\begin{figure}[t]
\setlength{\abovecaptionskip}{3pt}
\setlength{\belowcaptionskip}{0pt}
  \centering
  \includegraphics[scale=0.50]{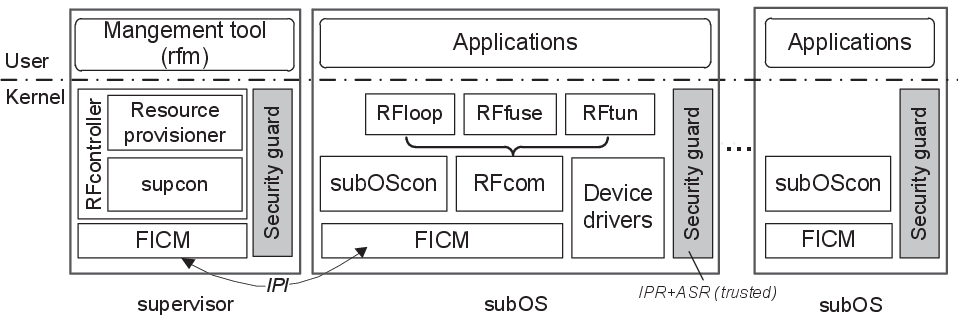}
  \caption{The structure of RainForest.}
  \label{architecture}
\end{figure}

Figure~\ref{architecture} depicts the RainForest structure. 
In our current implementation, the supervisor is the first  OS instance on each machine occupying at least one core or SMT thread and several MB of memory. 

  \emph{FICM} (\underline{F}ast \underline{I}nter-subOS \underline{C}ommunication \underline{M}eachnism) provides the basic interfaces for inter-subOS communication. FICM forks low-level message channels among subOSes based on inter-processor interrupt (IPI) and shared memory. A core is selected as the communication core in a subOS. On each subOS, we fork two FICM kernel threads (read/write) with real time priority to transfer \emph{tiny immediate messages} in units of cache lines (typically 64 Bytes) using NAPI  interfaces (New API, which combines interrupts and polling). It supports unicast, multicast, and broadcast operations. Upon FICM, \emph{supcon} in the supervisor and \emph{subOScon} in a subOS implement the primitives command of subOS management, respectively. Primitive commands issued from one end will be conducted and handled by the other. \emph{Resource provisioner} provides management interfaces for a full-fledged user space management tool---\emph{rfm}. Currently, the functions of \emph{Resource provisioner} and \emph{supcon} are implemented into a kernel module named \emph{RFcontroller}.

\subsection{SubOS management}\label{subos_management}


\begin{figure}[t]
\setlength{\abovecaptionskip}{3pt}
\setlength{\belowcaptionskip}{0pt}
  \centering
  \includegraphics[scale=0.50]{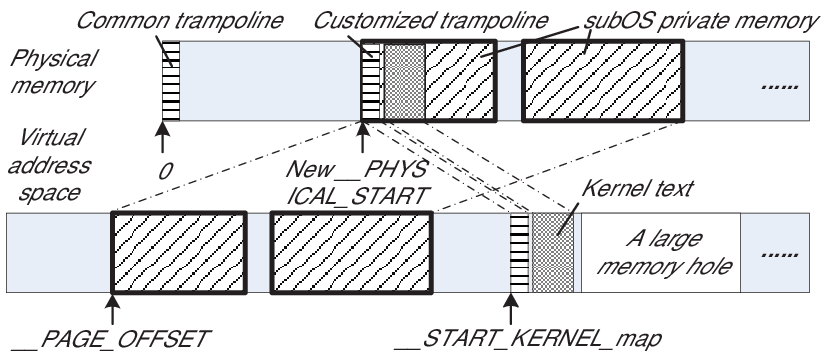}
  \caption{Physical memory and virtual memory organizations of a subOS.}
  \label{fig_bootup}
\end{figure}

\textbf{Memory organization}: Figure~\ref{fig_bootup} summarizes the organizations of physical and virtual memory spaces of a subOS. Conventionally, Linux directly maps all physical memory from zero to \emph{PAGE\_OFFSET} so that the conversion between virtual and physical addresses can be easily carried out by macros \emph{\_\_pa} and \emph{\_\_va}. In a subOS, we map the start address of its physical memory to \emph{PAGE\_OFFSET} and rephrase the \emph{\_\_pa} and \emph{\_\_va} macros. Other fix-mapped structures, which are mostly architecture-specific, are revised to create a safe booting environment. To support logical hot-add and hot-plug of memory regions whose physical addresses are ahead of the ones that are initially allocated to a subOS, we map them into a hole after \emph{\_\_START\_KERNEL\_map} in the address space instead of the direct-mapping address space starting from \emph{PAGE\_OFFSET}.



\textbf{Creating a subOS:} Before starting a subOS, RFcontroller prepares the least descriptions of hardware, upon which a subOS can successfully start up. In the normal Linux booting, it is actually conducted by BIOS. The information that RFcontroller fills for a subOS includes the description of SMP (MP Configuration Table), memory regions (E820 Table), and boot parameters. A new \emph{bootparam} parameter is added to instruct the kernel with a white list of passthrough PCIe endpoints. For the X86-64 architecture, we employ a two-phase jump after RFcontroller issues a restart command to the designated BSP (Bootstrap Processor) of a subOS. Once finishing the mode switch in the common trampoline in low memory, the execution will jump to the customized trampoline residing in its exclusive physical memory.


Though in RainForest each subOS independently manages its own physical resources, it still needs help from the supervisor. It is because a few hardware resources must be accessed via atomic operations or strictly protected from simultaneous updates. A typical example is the low memory ($<$ 1M) where resides a small slice of trampoline code that is needed for any X86 processor to finish mode switches.
Similar situations exist in the management of I/O APIC pins and PCI configuration registers. RainForest adopts global spinlocks stored in a globally shared page that is mapped into the address spaces of both the supervisor and subOSes.
Please note that this protection is specific to the x86-64 architecture.


\textbf{Resizing a subOS:}
Linux supports device hot-plug to allow failing hardware to be removed from a live system. However,  the shortest path of functioning or defunctioning a device for elastic resource adjustment in a subOS  is different from failing over hardware failures. We made great effort to shorten the path of logical hot-add and
hot-plug by removing several unnecessary phases. For example, we reduce the overhead of a CPU hot-add operation from ~150 ms to ~70 ms by removing the delay calibration of each CPU. For a memory hot-add operation, we reduce the scope of scanning removable page blocks through a short list recording the page blocks reserved by the kernel.

In RainForest, every resource adjustment operation is recorded by the supervisor, which can be further configured to collect the running information through subOScon from the \emph{proc} subsystem. In particular, hardware performance counters can be accurately employed using the \emph{perf} tools in a subOS to monitor architectural events of its CPUs. Besides, resource borrowing is allowed among subOSes but needs to be registered on the supervisor via a specified command.

\textbf{Destroying a subOS:} A subOS releases all resources before sending a shutdown request to the supervisor. The supervisor then prepares a designated trampoline program, forces it to transfer the CPUs to the supervisor via the trampoline, and finally reclaims other resources.

\subsection{Communication subsystems}\label{communication}
FICM is preferably used for light-weight messages.
To meet with different requirements of continuous and bulk communication, we introduce diverse communicating facilities including RFcom, RFloop, and other virtualization infrastructures.

RFcom exports high-level interfaces to kernel routines and user-space programs to facilitate inter-subOS communication. They are \emph{rf\_open}, \emph{rf\_close}, \emph{rf\_write}, and \emph{rf\_read}, \emph{rf\_map}, and \emph{rf\_unmap}. The former four interfaces operate on a socket-like channel, upon which subOSes easily communicate with packet messages. The other two interfaces help map  and unmap shared memory to individual address spaces, but without explicit synchronization mechanisms.

RFloop creates a fully-transparent inter-subOS network loop channel on the basis of Linux \emph{netfilter}. Network packets going to subOSes in the same machine will be intercepted on the network layer and transferred to the destination subOS. We achieve high bandwidth by adopting a lockless buffer ring and decreasing notification overhead using NAPI interfaces.

When the number of PCI-passthrough devices is insufficient for all subOSes, such as RAID controllers, RainForest adopts the splitting driver model applied in para-virtualization technologies. We base the network virtualization on the universal TUN/TAP device driver~\cite{tun_site} (RFtun), and the filesystem virtualization on FUSE~\cite{fuse_site} (RFfuse).
To gain better performance isolation and access control, we attach all back-end threads into a \emph{control group} created specifically for each subOS. Under the control group, resource consumptions of the back-end drivers can be accounted.

\subsection{State Sharing among the supervisor and subOSes}

Except for the mutually shared states in the private communication channels between two subOSes, the supervisor shares few states with all subOSes.

First, the supervisor manages a few globally-shared states that are publicly available to all subOSes. These states include the communication-core list, which is used for IPIs in FICM, and the MAC list, which is used to sniffer packets in RFLoop. The communication-core list is modified if the CPUs of a subOS change or a different load balancing algorithm is adopted for FICM. The MAC list is modified when a subOS is created or destroyed. As these updates are usually infrequent, state sharing can be implemented via either shared memory or messages.

Second, the supervisor reserves privileged operations on the protected
resources that can not be modified by a subOS. Typical examples include the low memory, the I/O APIC, and the PCIe configuration registers, as discussed in Section~\ref{subos_management}.

\subsection{Refactoring Efforts}\label{refactoring}

Most of the Linux software stack is unmodified to support the existing Linux ABI. Table~\ref{refactoring_efforts} summarizes  the new and modified lines of code in RainForest, most of which are performed in the portable functions and independent modules. Out of the key effort in the kernel change, the largest portions are FICM and the functionalities of the supervisor, which can be dynamically activated or deactivated.


%
%

\begin{table}[h]
\renewcommand{\arraystretch}{1.1}
\setlength{\abovecaptionskip}{3pt}
\setlength{\belowcaptionskip}{0pt}
\setlength{\textfloatsep} {0pt plus 2pt}
\setlength{\tabcolsep}{5pt}
\caption{The effort of building RainForest on the basis of a Linux 2.6.32 kernel.}
\centering
\begin{lrbox}{\tablebox}
\begin{tabular}{l|c}
 \hline
 \textbf{Component} & \textbf{Number of Lines}\\ \hline
 The supervisor & 994 \\
 The RFcontroller module & 1837 \\
 A subOS & 1969 \\
 FICM / RFcom / subOScon &  1438 / 1522 / 1331 \\
 RFloop / RFtun / RFfuse & 752 / 2339 / 980 \\
 others (rfm) & 4789\\
 \hline
\end{tabular}
\label{refactoring_efforts}
\end{lrbox}
\scalebox{0.8}{\usebox{\tablebox}}
\end{table}

\section{Evaluation}\label{section_evaluation}
RainForest currently runs well on x86-64 platforms (i.e., Intel Xeon E5620, E5645, E5-2620, E5-2640, and E7-8870). We run the benchmarks on the two servers listed in Table~\ref{table_server_config}. In fact, we have tested two benchmark suites (lmbench~\cite{staelin2002lmbench3} and PARSEC~\cite{bienia2008parsec}) to demonstrate the zero overhead of utilizing the IFTS OS model relative to the virtualization techniques of LXC and Xen. The results are consistent with those reported in ~\cite{soltesz2007container} and we do not represent here. Other comprehensive comparisons are performed on the three Linux kernels, LXC, Xen and IX.

The special technologies including Intel Hyper-Threading, Turbo Boost, and SpeedStep are disabled to ensure better fairness.
We build all the systems (Linux, containers in LXC, Dom0 and DomU in Xen, and subOSes in RainForest) on \emph{SUSE Linux Enterprise Server 11 Service Pack 1}, which is released with the 2.6.32 kernel. Using the original distribution, we compile kernels \emph{3.17.4}, \emph{4.9.40}, and \emph{2.6.35M} so as to compare with  state-of-the-practice. The Linux and Xen versions are 0.7.5 and 4.0.0, respectively. \emph{In this Section, we use \emph{OS instance} to denote a container of LXC, a DomU guest OS on Xen, or a subOS in RainForest.} For RainForest, as the operation overhead of the supervisor is negligible, we also deploy applications on it in our experiments. This ensures the full utilization of the machine resources.
For LXC and Xen, we bind each OS instance's CPUs or virtual CPUs to physical CPUs, which helps decrease the influence of process scheduling. The NUMA memory allocation policy is \emph{local allocation}, which preferably allocates \emph{local} memory in the same node to their CPUs accordingly. For the Linux kernels, we do not pin the benchmarks on specific CPUs or specify the memory allocation policies, making them compete for the system resources without restriction. Since there are abundant ethernet adapters in a single server, we make each OS instance directly access one to get near-native performance. For the storage, we attach each OS instance a directly accessed disk when deploying two OS instances in a 12-core server, one under an LSI SAS1068E SAS controller and the other under an Intel ICH10 AHCI controller.
When running four OS instances on a \emph{S-B}-type server, we adopt the \emph{tmpfs} in-memory file systems to reduce the contention for a single disk.

\begin{figure}[t]
\setlength{\abovecaptionskip}{0pt}
\setlength{\belowcaptionskip}{0pt}
  \centering
    \subfloat[malloc]{\label{fig_root_cause_malloc_cdf}\includegraphics[width=0.33\linewidth]{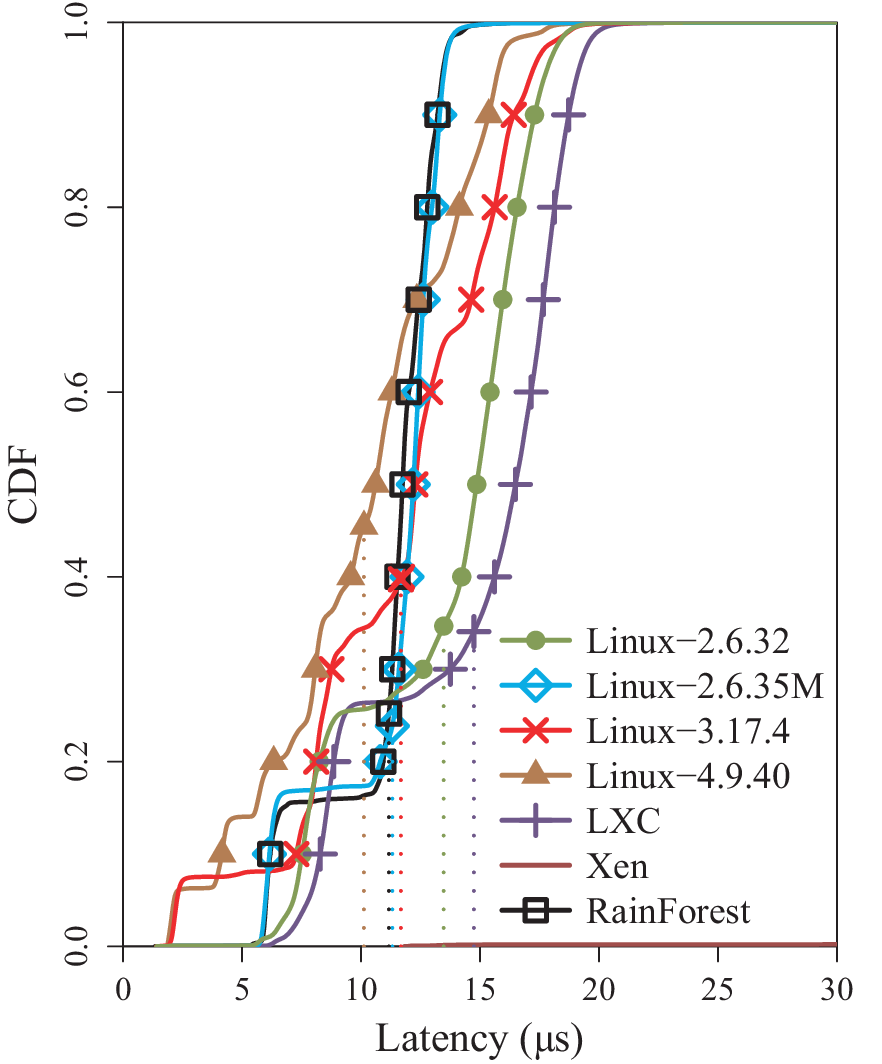}}
    \subfloat[munmap]{\label{fig_root_cause_munmap_cdf}\includegraphics[width=0.33\linewidth]{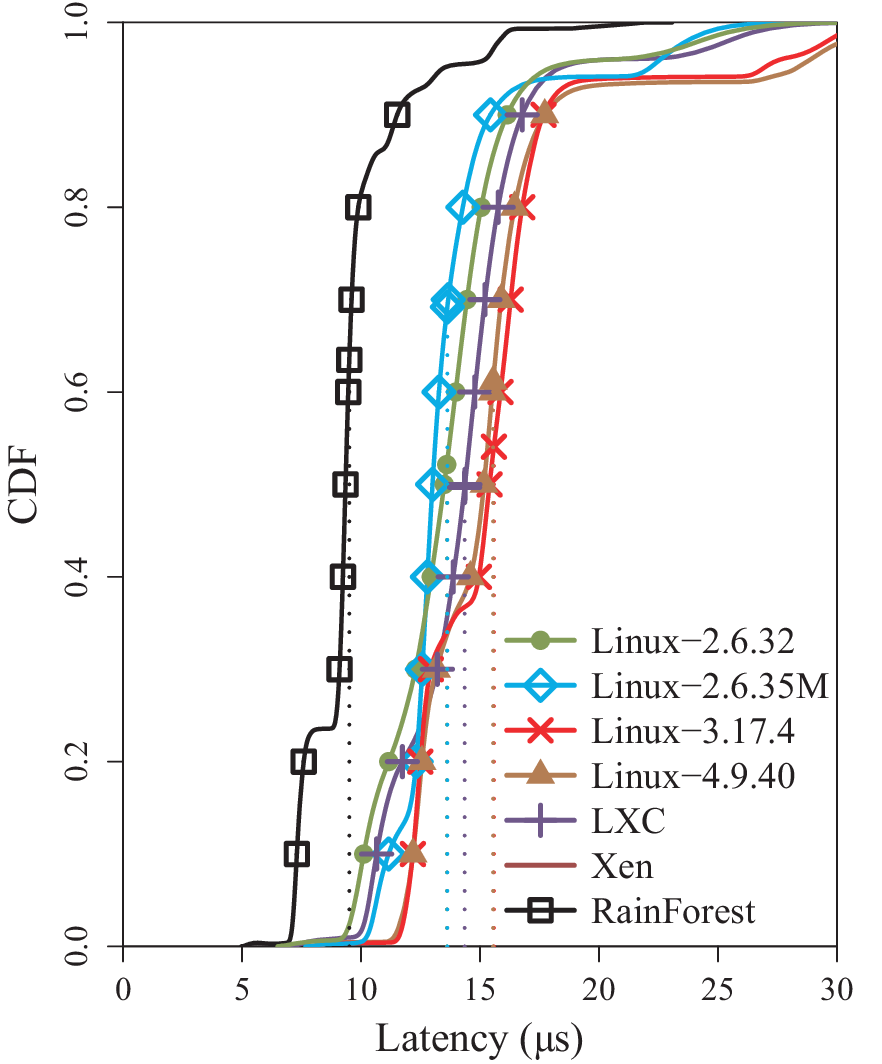}}
    \subfloat[open]{\label{fig_root_cause_open_cdf}\includegraphics[width=0.33\linewidth]{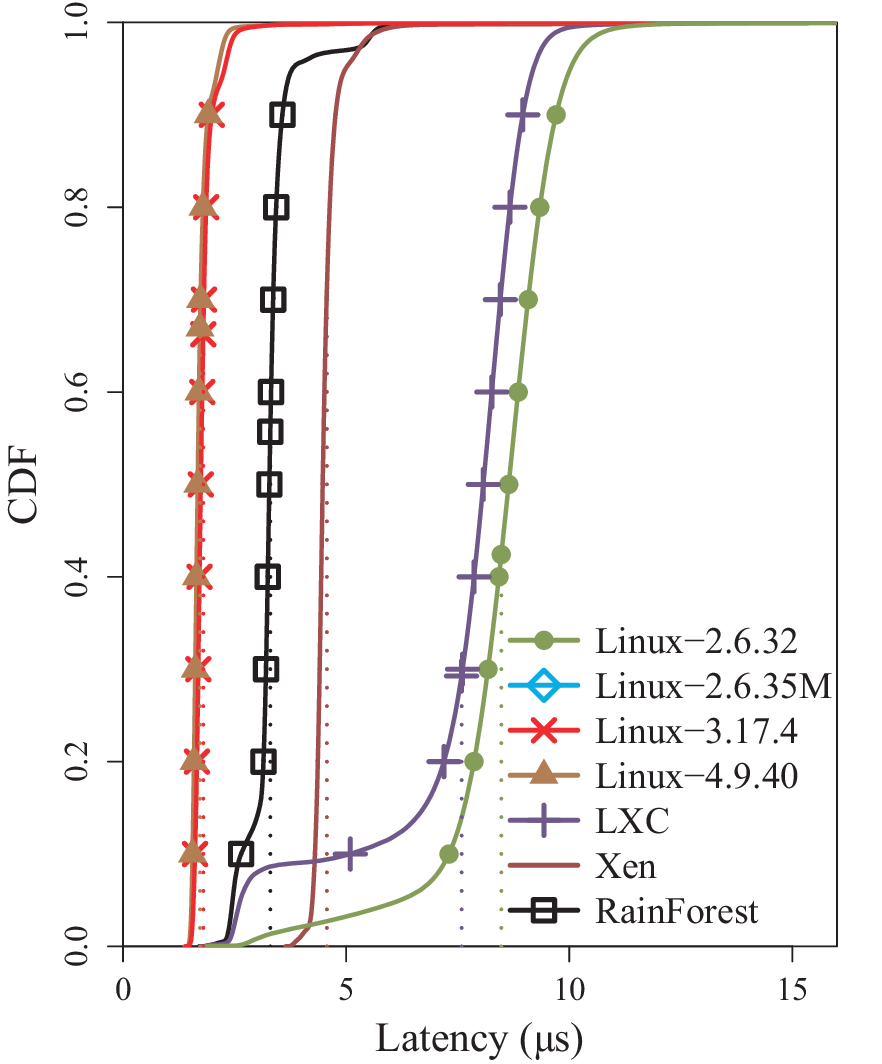}}
    \caption{The cumulative latency distribution of three microbenchmarks. \emph{The number on the system with the maximum tail latency is not reported in each subFigure.}}
    \label{fig_rootcause_microbenchmarks_cdf}
\end{figure}

\subsection{Understanding the root causes}\label{hit_root_cause}

As an application benchmark is about several
orders of magnitude complex than a micro-benchmark, it is difficult to run application benchmarks to understand the OS kernel behaviours. 
In stead, we run the Will-It-Scale benchmark suite to inspect the advantage of the IFTS architecture and uncover the root causes mentioned in Section~\ref{motivation_discussion}. We deploy six OS instances on an S-A type server. Each OS instance runs a process and the number of threads increases with the cores. 

 We deliberately  only present the performance numbers of three benchmarks:  $malloc$, $munmap$, and $open$. Among them, the $malloc$ operations contend for the free list locks~\cite{Michael:2004:SLD}; the $munmap$ operations cause TLB shootdown to other CPUs~\cite{DiDi2011}; the $open$ operations competes for both $locks$ and $dentry$ cache~\cite{Tsai:2015:GMV}. Note that RainForest does not intend to eliminate the causes (3) and (5)  in Section~\ref{motivation_discussion}, as well as other systems.

Figure~\ref{fig_rootcause_microbenchmarks_cdf} shows the cumulative latency distribution of each benchmark. The dotted vertical lines indicate the average latencies. Except for $open$, RainForest exhibits the best performance in terms of both the  maximum latencies and standard deviations.
Excluding Linux-3.17.4 and Linux-4.9.40, we also notice that RainForest exhibits lower tail latencies and average latencies in all the benchmarks.
For $malloc$, Linux-4.9.40 has the lowest average latency, but it has higher  maximum latency and standard deviation with respect to RainForest. For $open$ both Linux-3.17.4 and Linux-4.9.40 shows better tail, average, and maximum latencies with respect to RainForest.
Please note that RainForest is based on Linux 2.6.32, and the performance improvement of Linux kernel evolution is definitely helpful to RainForest.

\subsection{Performance isolation}
In this section, we use both micro benchmarks and application benchmarks to measure (average or worst-case) performance isolation of different systems on an \emph{S-A} type server with two OS instances.

\subsubsection{Mixed microbenchmarks}

The micro benchmarks we used include SPEC CPU 2006~\cite{henning:2006:speccpu} (CPU intensive), cachebench~\cite{Mucci:1998:cachebench} (memory intensive), netperf~\cite{jones:1996:netperf} (network intensive), IOzone~\cite{iozone_site}(filesystem I/O intensive). 11 benchmarks from four benchmark suites are selected to exert heavy pressures to different subsystems. We investigate the mutual interference of any two benchmarks which are deployed in two OS instances in a single server. The two OS instances, each with 3 cores and 8 GB memory, run in the same NUMA node rather in two NUMA nodes, producing more performance interference. As the processes contend heavily on accessing (especially writing) files on a disk in the IOzone case, we make them read/write files in the tmpfs file system to stress the filesystem caches and memory instead of physical disks. For cachebench, we set the footprint to be 32 MB, which is greater than the LLC and makes more pressure on the memory system. When running two netperf benchmarks on Linux, each of them uses a NIC with an IP in different networks.

\begin{figure*}[t]
\setlength{\abovecaptionskip}{3pt}
\setlength{\belowcaptionskip}{0pt}
  \centering
    \subfloat[Linux-3.17.4]{\label{fig_isolation.2285.micro.Linux.3.17.4}\includegraphics[width=0.24\linewidth]{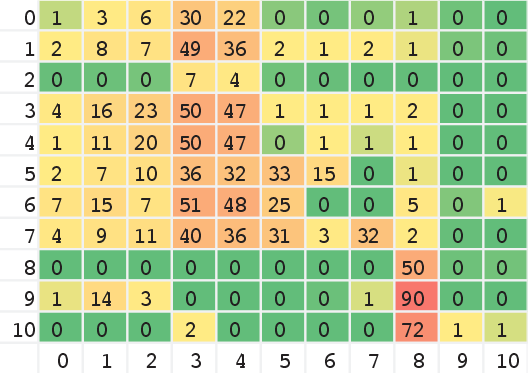}}
    \subfloat[LXC]{\label{fig_isolation.2285.micro.LXC}\includegraphics[width=0.24\linewidth]{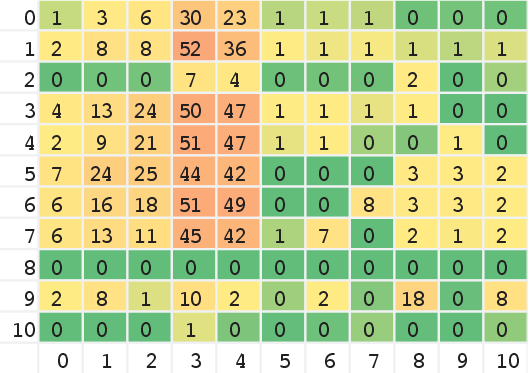}}
    \subfloat[Xen]{\label{fig_isolation.2285.micro.Xen}\includegraphics[width=0.24\linewidth]{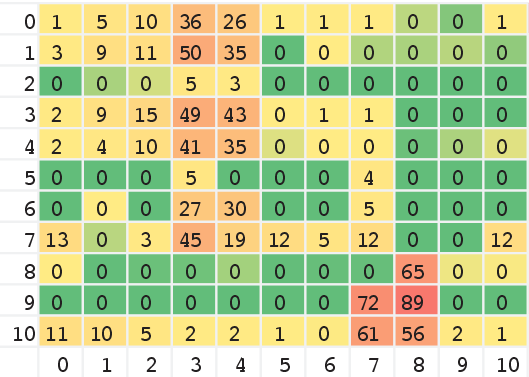}}
    \subfloat[RainForest]{\label{fig_isolation.2285.micro.RF}\includegraphics[width=0.24\linewidth]{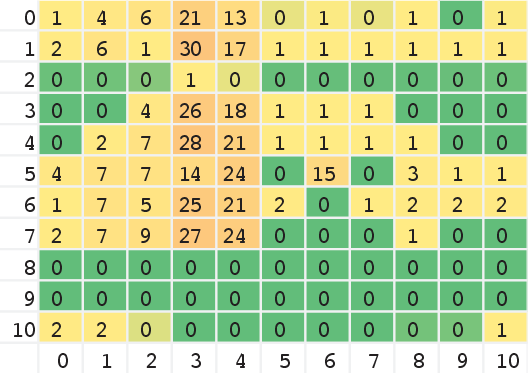}}
    \caption{Performance degradation of co-running two benchmarks on a single server. Numbers 0$\sim$10 on the both \emph{x-} and \emph{y-} axes denotes \emph{SPECCPU.\{bzip2, sphix3\}}, \emph{cachebench.\{read, write, modify\}}, \emph{IOzone.\{write, read, modify\}}, \emph{netperf.\{tcp\_stream, tcp\_rr, tcp\_crr\}}, respectively. The numbers in the grid are the average performance slow down percentages (\%) when a foreground benchmark on y-axis interfered by a background one on x-axis relative to solo-running a foreground one.}
    \label{fig_isolation_microbenchmarks}
\end{figure*}

Figure~\ref{fig_isolation_microbenchmarks} reports the \emph{average} performance degradation of each foreground benchmark affected by the background one with respect to solo running the foreground one. RainForest exhibits good performance isolation except for running cachebench.write and cachebench.modify benchmarks as backgrounds. LXC and Xen have heavier performance interference in the same two scenarios, and the performance is degraded much in the other scenarios. Besides, different Linux kernels show similar behaviors (only report Linux-3.17.4), indicating the performance isolation is poor in an SFTI OS architecture as it shares many data structures  protected by locks.


\subsubsection{Improving worst-case performance}
In this section, we test and verify RainForest's ability of improving the worst-case performance and resource utilization. The latency-critical workload we choose is \emph{Search} from BigDataBench~\cite{Wang:2014:BigDataBench}. The front-end Tomcat server distributes each request from the clients to all back-end servers, merges the records received from the back ends, and finally responds to the clients. Tomcat is not the bottleneck according to our massive tests. In the experiments, we choose the real workload trace and set the distribution to be uniform---sending  requests at a uniform rate.

\begin{figure}[t]
\setlength{\abovecaptionskip}{3pt}
\setlength{\belowcaptionskip}{0pt}
  \centering
  \includegraphics[scale=0.45]{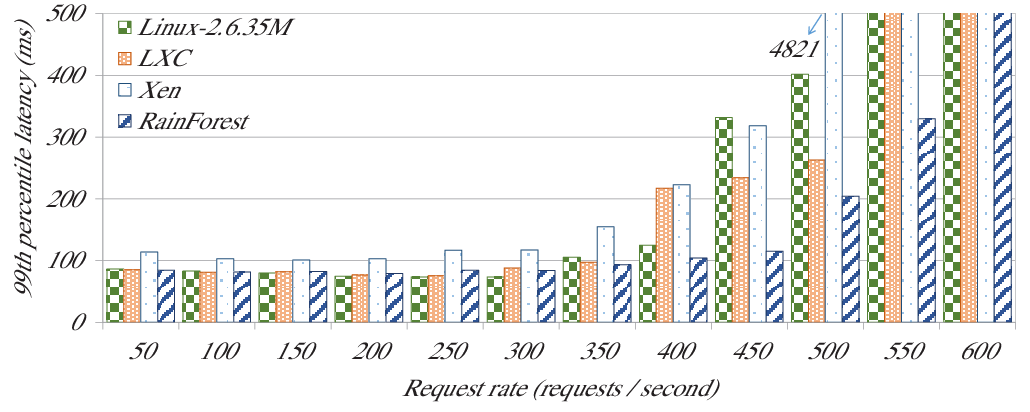}
  \caption{Tail latency  of Search running on two OS instances on a single server.}
  \label{tail_latency_gurantee_search_99latency}
\end{figure}

When running a single Search workload, we found the tail latency dramatically climbs up to seconds when the request rate reaches 400 req/s, while the CPU utilization rate cannot surpass 50\%. So we set up two Search backends on two OS instances on the same server, each with 6 cores and 16 GB memory. Figure~\ref{tail_latency_gurantee_search_99latency} illustrates the tail latencies with increasing load levels. For three Linux kernels, we only show Linux-2.6.35M as it exhibits better worst-case performance. When the load level increases to 400 requests/s, the tail latencies on LXC and Xen deteriorate significantly. The tail latencies are beyond 200 ms except for RainForest. Linux  still keeps tail latency below 200 ms owing to free scheduling across all 12 cores, but the tail latency gets much worse after 450 requests/s, indicating  aggressive resource sharing will  produce more interference. If we demand the endurable limit of the tail latency is 200ms, the maximum throughput of Linux 2.6.35M, LXC, Xen, and RainForest is around  400, 350, 350, 500 request/s, respectively. On these load levels, the CPU utilization is 59.8\%, 58.0\%, 55.7\%, and 69.7\%, respectively. Although the CPU utilization of Linux can finally reach to $\sim$90\% at 600 req/s, the tail latency becomes totally unacceptable.




\subsubsection{Services mixed with offline workloads}\label{mixed_search_parsec_static}
\textrm{\\} 
Co-running online services and offline batches is always a thorny problem.
We investigate RainForest's performance isolation in terms of co-running Search with batch workloads. In this test, each OS instance runs on 6 cores and 16 GB memory within a NUMA node. We run Search and a PARSEC workload on each OS instance, respectively. The baseline is solo running Search on one OS instance of each system. The requests are replayed in a uniform distribution at 300 requests/s, beyond which the Search tail latency on  Xen will deteriorate dramatically.

\begin{figure}[t]
\setlength{\abovecaptionskip}{3pt}
\setlength{\belowcaptionskip}{0pt}
  \centering
  \includegraphics[scale=0.45]{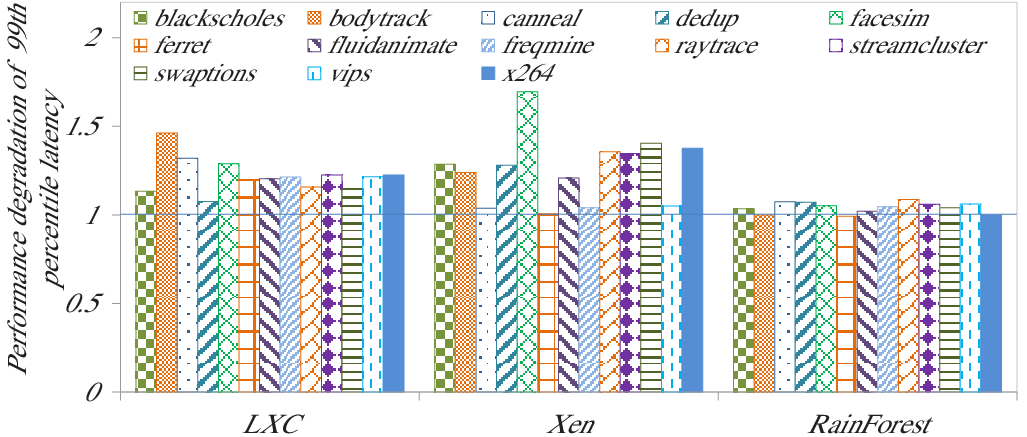}  \caption{Tail latency slowdown of Search when co-locating with Parsec benchmarks relative to solo running. The Search tail latency when solo running at 300 req/s  on LXC, Xen, and RainForest is 129.3, 210.9, and 128.5 ms, respectively. The performance numbers of Linux kernels are shown in Figure~\ref{fig_motivation_tail_latency_search}.}
  \label{fig_tail_latency_parsec}
\end{figure}

Figure~\ref{fig_tail_latency_parsec} illustrates the performance degradation of the Search tail latency with respect to the baseline. As shown in  Figure~\ref{tail_latency_gurantee_search_99latency}, Xen has the poorest tail latency performance (210.9 ms) when running Search at 300 req/s. Actually, even at 250 req/s, its tail latency is high as 150.6 ms, worse than LXC and RainForest, and the average slowdown reaches up to 25.6\%. As the virtualization overhead exists, Xen is not suited for a high load level when the tail latency matters, and its performance is worse than LXC.

RainForest exhibits good performance isolation almost in all the cases, while the other systems get larger tail latencies in many cases. The slowdown of tail latency in RainForest is always less than 8\%, while that of LXC is high as 46\%. If we only care about the average performance of offline batches~\cite{barham2003xen}, Xen gains better performance than Linux. But unfortunately, in this case, the tail latency of Search deteriorates worse, which is totally unacceptable.




\begin{table}[t]
\renewcommand{\arraystretch}{1.1}
\setlength{\abovecaptionskip}{3pt}
\setlength{\belowcaptionskip}{0pt}
\setlength{\textfloatsep} {0pt plus 2pt}
\setlength{\tabcolsep}{5pt}
\centering
\label{table_elastic_partition}
\caption{Overhead of adjusting three systems (in seconds).}
\label{table_elastic_partition}
\begin{lrbox}{\tablebox}
\begin{tabular}{|c|c|c|c|c|c|c|}
\hline
Configuration & \multicolumn{2}{c|}{6 CPUs, 16G RAM} & \multicolumn{2}{c|}{1 CPU} & \multicolumn{2}{c|}{512M RAM} \\ \hline
Operations & create & destroy & online & offline & online & offline \\ \hline
LXC & 2.1 & ~0 & 0.002 & 0.002 & 0.002 & 0.002 \\ \hline
Xen & 14.2 & 5.9 & 0.126 & 0.127 & 0.167 & 0.166 \\ \hline
RainForest & 6.1 & ~0 & 0.066 & 0.054 & 0.020 & 0.060 \\ \hline
\end{tabular}
\end{lrbox}
\scalebox{0.8}{\usebox{\tablebox}}
\end{table}

\subsection{Flexibility of resource sharing}
In this section, we evaluate the overhead and agility of adjusting resources when consolidating Search with varying other workloads.

\subsubsection{Evaluating elasticity overhead}
\textrm{\\} 
We evaluate the overhead of creating, destroying, and resizing the OS instance  by  performing these operations 100 times on an \emph{S-A} type server. For two OS instance configurations, Table~\ref{table_elastic_partition} lists the overheads on LXC, Xen, and RainForest.

From the experimental results, we can find that the elasticity of these systems can be overall described as LXC $>$ RainForest $>$ Xen. For LXC, allocating and deallocating  resources are not really conducted on physical resources but performed by updating the filter parameters of the cgroup subsystem. Although an adjustment operation is always quickly responded, it may take a long time to take effect. For Xen, the adjusting phase is also longer. Besides getting emulated resources ready before loading a domU kernel, VMM needs to prepare several software facilities, including setting up shadow page tables, shared pages, communication channels, etc.

\begin{figure}[t]
\setlength{\abovecaptionskip}{3pt}
\setlength{\belowcaptionskip}{0pt}
  \centering
  \includegraphics[scale=0.43]{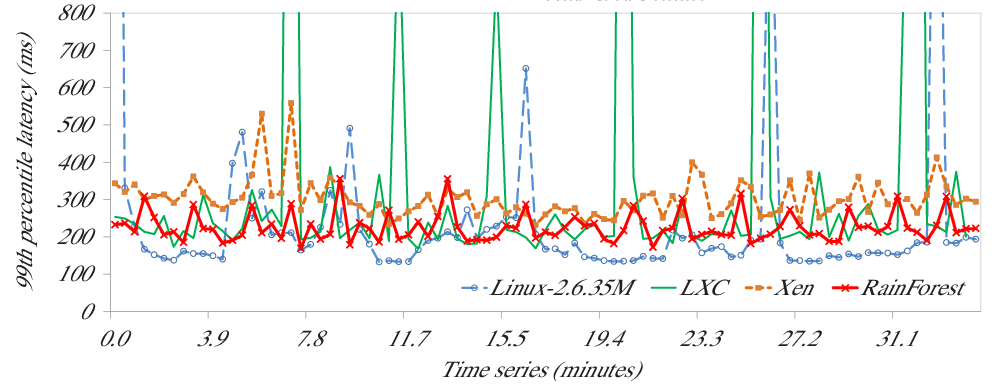}
  \caption{Search tail latency varies with time. The threshold $(lt, ut)$ is (160, 200). For Linux, only Linux 2.6.35M is shown as it gains better performance than Linux 2.6.32 and Linux 3.17.4.}
  \label{fig_tail_latency_consolidation_parsec}
\end{figure}

\begin{figure}[t]
\setlength{\abovecaptionskip}{3pt}
\setlength{\belowcaptionskip}{0pt}
  \centering
  \includegraphics[scale=0.43]{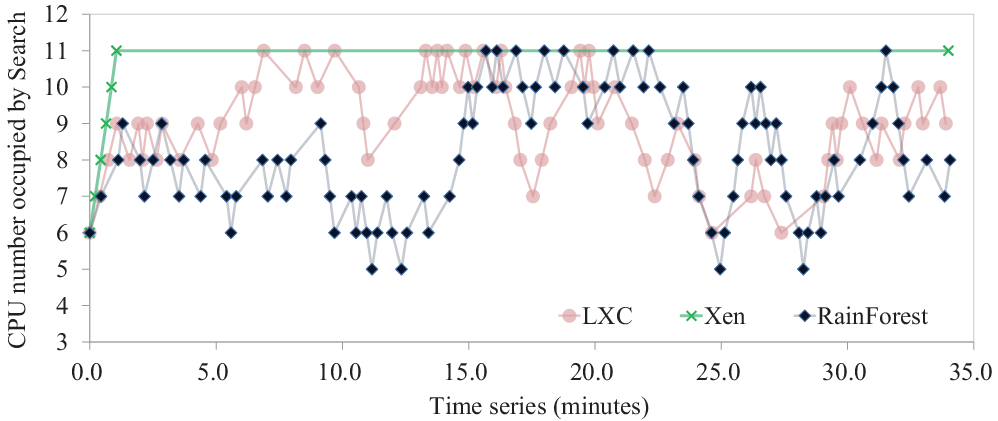}
  \caption{The varying number of the CPU owned by the OS instance running Search v.s. time. The tail latency threshold  $(lt, ut)$ is set to be (160, 200).
}
  \label{fig_cpu_number_of_nutch_consolidation_parsec}
\end{figure}



\begin{table}[t]
\renewcommand{\arraystretch}{1.1}
\setlength{\abovecaptionskip}{3pt}
\setlength{\belowcaptionskip}{0pt}
\setlength{\textfloatsep} {0pt plus 2pt}
\setlength{\tabcolsep}{5pt}
\centering
\caption{Performance of consolidating Search and PARSEC benchmarks. We identify requests invalid if the response time $>$ 1 second and exclude them from the throughput.}
\label{table_consolidation_results}
\begin{lrbox}{\tablebox}
\begin{tabular}{|c|c|c|c|c|}
\hline
 & PARSEC batch & \multicolumn{2}{c|}{Search} \\ \hline
Platform &  running time & 99\%th tail & throughput \\ \hline
Units & seconds & ms & req/s \\ \hline
Linux-2.6.32 & 1058.7 & 379.3 & 217.1\\ \hline
Linux-2.6.35M & 1136.5 & 371.4 & 217.9  \\ \hline
Linux-3.17.4 & 1054.4 & 410.8  & 219.1 \\ \hline
LXC & 1716.8 & 284.1 & 214.0 \\ \hline
Xen & 4731.0 & 305.4 & 209.4\\ \hline
RainForest& 1520.0 & 230.2 & 214.4\\ \hline
\end{tabular}
\end{lrbox}
\scalebox{0.8}{\usebox{\tablebox}}
\end{table}

\subsubsection{Agile changes to dynamic workloads}\label{agility}
\textrm{\\} 
Here we evaluate the agility of the three systems to fast-changing workloads.
We initially configure the server with two OS instances as in Section~\ref{mixed_search_parsec_static}. But the request rate is fluctuated according to the original distribution. We \emph{package 13 PARSEC benchmarks into a batch job} and run it repeatedly in one OS instance while Search runs in the other OS instance. Meanwhile, a simple scheduler is adopted to adjust the resources of the two OS instances to reduce the tail latency of Search. In this test, \emph{we only adjust the CPUs of two OS instances} rather than finding the optimal strategy of adjusting all resources, which is an open issue. We set two thresholds $(lt, ut)$ to bound the tail latency. That is if the tail latency of the last 10 seconds is above the upper threshold---$ut$, a CPU will be transferred from the PARSEC OS instance to the other, and vice versa.

We record the CPU adjustment events and tail latency variations throughout the replay of 482400 requests in 37.5 minutes. From Figure ~\ref{fig_tail_latency_consolidation_parsec}, we observe that both Linux and LXC have large fluctuations, showing unstable worst-case performance. Xen has smaller fluctuations but the tail latency even exceeds 500 ms. Relatively, RainForest exhibits the most stable worst-case performance, mostly between 200 ms and 300 ms. Meanwhile, after replaying all the requests, the number of PARSEC benchmarks finished on Linux-2.6.32, Linux-2.6.35M, Linux-3.17.4, LXC, Xen, and RainForest is  22, 26, 25, 19, 6, and 20,  respectively. Xen finished less PARSEC benchmarks because the tail latency cannot be reduced to below 200 ms even 11 cores are occupied. RainForest outperforms the other systems in terms of both the worst-case performance of Search and the average  performance of the batch jobs in terms of running time.
Figure~\ref{fig_cpu_number_of_nutch_consolidation_parsec}  records the varying number of the processors owned by the OS
instance when running Search. In Linux, Search and PARSEC workloads compete for resources adversely and we do not report the number in Figure~\ref{fig_cpu_number_of_nutch_consolidation_parsec}.

%
Table~\ref{table_consolidation_results} reports the running time of the PARSEC batch jobs and the corresponding performance of Search on each system. The Search throughput on each system  does not differ significantly, while the 99\% tail latencies are quite different. In RainForest, the tail latency is the lowest (230.2 ms) and the batch job runs faster (1520.0 seconds) than LXC and Xen. Interestingly, we also observe Linux 3.17.4 gains the best average performance in terms of the running time of the PARSEC benchmarks and the Search throughput, however, its tail latency is the worst (high as 410.8). It again confirms that the SFTI OS architecture is optimized toward the average performance. We do not test Linux 4.9.40 because of time limitation in the rest of experiments.

\subsection{Scalability}
We initiate  four OS instances for LXC, Xen, and RainForest on a \emph{S-B} type server. Each OS instance has 10 cores and 250GB RAM, among which 50GB is used for the tmpfs file system.

\begin{figure}[t]
\setlength{\abovecaptionskip}{3pt}
\setlength{\belowcaptionskip}{0pt}
  \centering
    \subfloat[99th percentile latency]{\label{scalability.5885.memcached.multi-process}\includegraphics[width=0.48\linewidth]{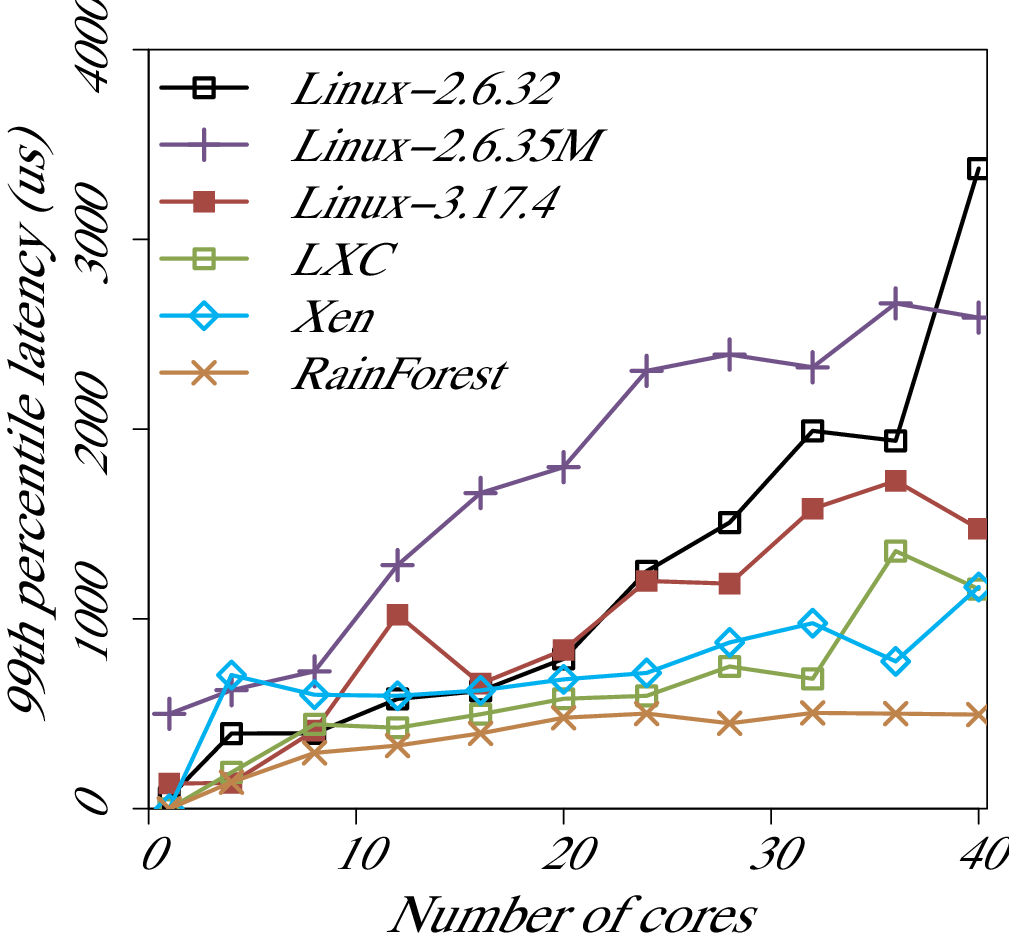}}
    \subfloat[Throughput]{\label{scalability.5885.memcached.multi-thread}\includegraphics[width=0.48\linewidth]{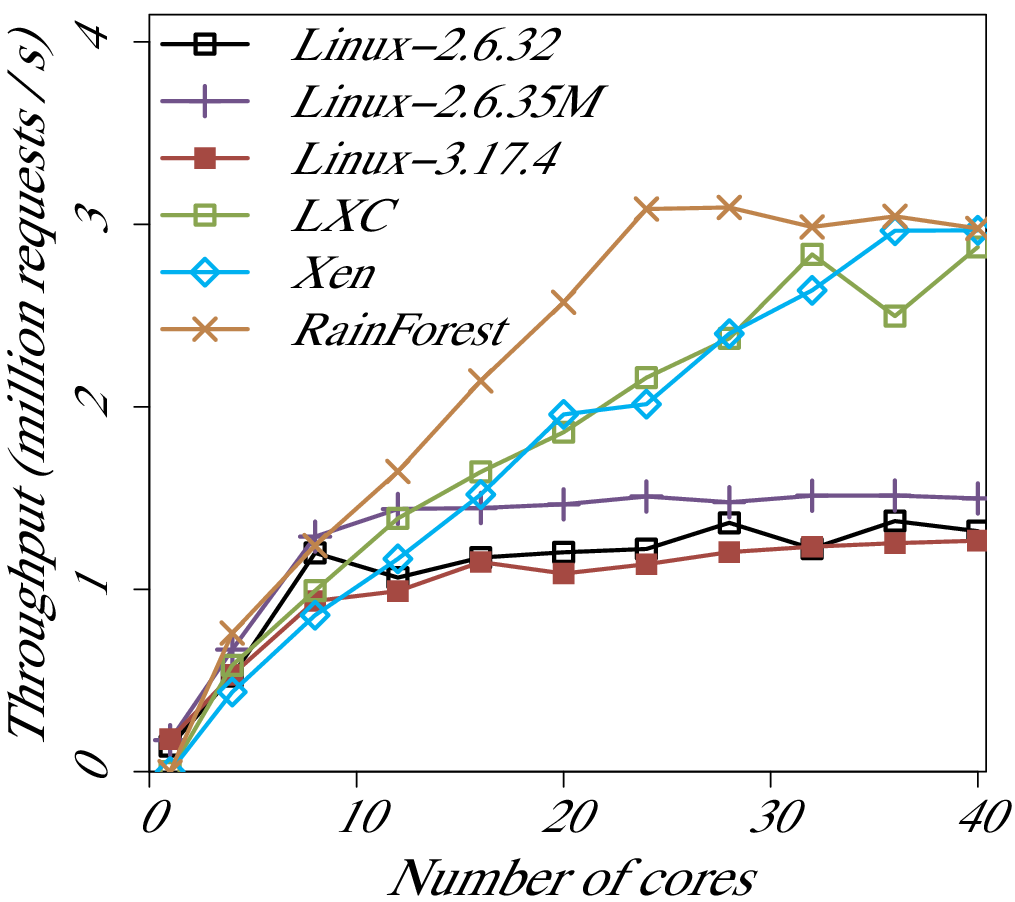}}
    \caption{Scalability in terms of tail latency running memcached on different systems.} 
    \label{fig_scalability.5885.memcached}
\end{figure}

\subsubsection{Scalability in terms of worst-case performance}
\textrm{\\} 
We use an in-memory key-value store (memcached) to evaluate the scalability in terms of the tail latency. The memcached benchmark we use is a variant of that in MOSBench~\cite{Boyd-Wickizer:2010:MOSBench}. Similar to the method in Section~\ref{Section_Motivation}, we run multiple memcached servers in Linux, each on its own port and being bound to a dedicated core. For the other systems, the requests are sent averagely into four OS instances. We profile the lookup time of all requests and present the tail latencies with increasing cores in Figure~\ref{fig_scalability.5885.memcached}. Although the scalability problem still exists, the tail latency on RainForest increases slowly than the others. When the core number is 40, the tail latency improvements of RainForest in comparison with Linux 2.6.32, Linux 2.6.35M, Linux 3.17.4, LXC, and Xen are 7.8x, 4.2x, 2.0x, 1.3x, and 1.4x, respectively. Among the three Linux kernels, Linux 2.6.35M gets the highest throughput. 
RainForest gains the highest throughput among the six systems.

%
%

\begin{figure}[t]
\setlength{\abovecaptionskip}{3pt}
\setlength{\belowcaptionskip}{0pt}
  \centering
  \includegraphics[scale=0.43]{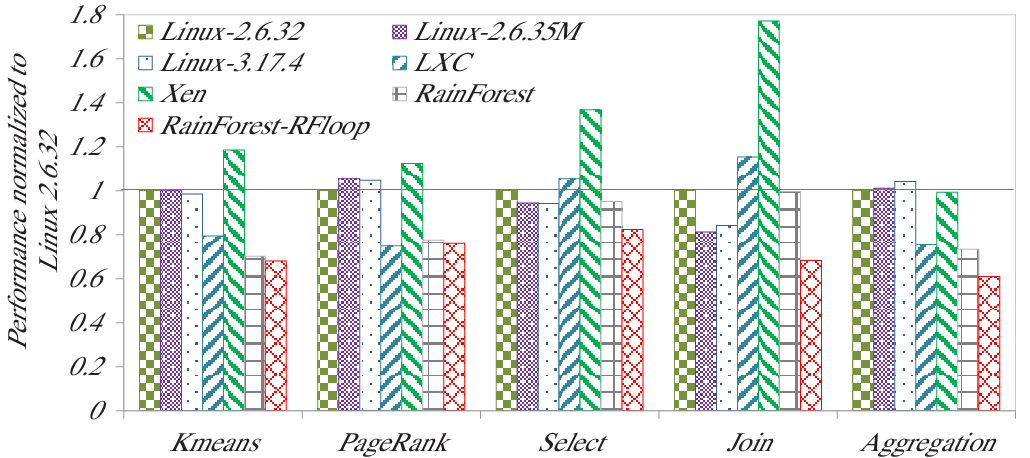}
  \caption{Spark workloads performance.} 
  \label{fig_spark_all_workloads}
\end{figure}

\subsubsection{The average performance}\label{evaluation_spark}
\textrm{\\} 
 On Linux, we use the "standalone" deploy mode with the  4 worker instances on LXC, Xen, and RainForest (each OS instance runs a worker instance). The Spark~\cite{zaharia2012resilient} workloads include OLAP SQL queries (Select, Aggregation, and Join) from~\cite{Wang:2014:BigDataBench, pavlo2009comparison} and offline analytic algorithms (Kmeans and PageRank).

Figure~\ref{fig_spark_all_workloads} shows their performance on the systems, including RainForest with RFloop enabled. Although the bandwidth limit of the memory controllers and peripherals may be bottlenecks, we also get much improvement from RainForest. The maximum speedup is 1.43x, 1.16x, and 1.78x  compared to Linux, LXC, and Xen, respectively. With RFloop enabled, the maximum speedup is  1.71x, 1.69x, and 2.60x, respectively. For Join and Aggregation, the time consumed in data shuffling among Spark workers takes a high proportion out of the whole execution. RFloop facilitates them with fast communication channels. Using netperf, the TCP stream performance of RFloop is 0.63x of the local loop, 1.47x of a virtual NIC of Xen, and 15.95x of a physical NIC. For the UDP stream test, the speedup is 0.44x, 13.02x, and 4.34x, respectively.

As many workloads still benefit from shared memory, the number of OS instances where the application is distributed influences the performance. We tested Select, Aggregation, and Join on 2, 4, 6, and 8 subOSes using the same server. We find that the optimal subOS number is four for Select and Aggregation, while for Join eight subOSes gets the best performance (the improvement can achieve 170\%).

\section{Experience and Limitations}






RainForest is a concrete implementation of our model. However, it is only a point in the implementation space.

First, implementing a subOS as a Linux-like OS instance is not required by the model. However, it is commercially significant~\cite{banga1999resource} for DC workloads.  

Second, we have compared RainForest with Barrelfish using the $munmap$ microbenchmark (we failed to run application benchmarks). Barrelfish shows good scalability consistent with ~\cite{ Baumann:2009:multikernel}. But with increasing memory size to be unmapped, the average and tail latencies deteriorate significantly. The reason might reside in the two-phase commit protocol, used to ensure global consistence. In this case, the message passing system needs to queue large amount of messages caused by splitted unmap operations, indicating the potential bottleneck for the message passing approach.

Third, as the supervisor and  subOSes run at the same privileged level, our architecture can be easily extended to multiple coherence domains that have no hardware cache coherence. However, its flexibility of resource sharing will be limited and we can only elastically partition resources within a domain, or else we have to leverage the DSM technology among several domains in ~\cite{lin2014k2}.

Finally, the number of subOSes is strictly limited by the cores or SMT threads, but we can  integrate containers or guest OSes in a subOS. Two-level scheduling is  an interesting open issue.
When the peripheral devices are not enough or sharing them may not significantly slow down the application performance on other subOSes, we can leverage the virtualization technology like the split-driver model~\cite{chisnall2008definitive, liu2006high}.

\section{Conclusions}

In this paper, we propose an IFTS OS model, and explore the space of applying the IFTS OS model to a concrete OS implementation:  RainForest, which is  fully compatible with Linux ABI. Our comprehensive evaluations  demonstrate RainForest outperforms Linux with four kernels, LXC, and Xen in terms of worst-case and average performance most of time when running a large number of benchmarks. 

\bibliographystyle{plain}
\bibliography{RainForest}

\begin{thebibliography}{100}

\bibitem{iozone_site}
{IOzone}. {Accessed Dec 2014.}
\newblock \url{http://www.iozone.org/}.

\bibitem{PARSEC_Source_code}
{PARSEC 3.0 benchmark suite. Accessed July 2014.}
\newblock \url{http://parsec.cs.princeton.edu}.

\bibitem{SUN_Dynamic_system_domains}
{Sun Enterprise 10000 Server: Dynamic System Domains. Accessed July 2014.}
\newblock
  \url{http://docs.oracle.com/cd/E19065-01/servers.e25k/817-4136-13/2_Domains.html
  }.

\bibitem{Will-It-Scale}
{Will-It-Scale benchmark suite. Accessed July 2014.}
\newblock \url{https://github.com/antonblanchard/will-it-scale}.

\bibitem{agmon2012resource}
Orna Agmon Ben-Yehuda, Muli Ben-Yehuda, Assaf Schuster, and Dan Tsafrir.
\newblock The resource-as-a-service (raas) cloud.
\newblock In {\em Proceedings of the 4th USENIX conference on Hot Topics in
  Cloud Ccomputing}, pages 12--12. USENIX Association, 2012.

\bibitem{Ahn:2016:IIR}
Sungyong Ahn, Kwanghyun La, and Jihong Kim.
\newblock Improving i/o resource sharing of linux cgroup for nvme ssds on
  multi-core systems.
\newblock In {\em Proceedings of the 8th USENIX Conference on Hot Topics in
  Storage and File Systems}, HotStorage'16, pages 111--115, Berkeley, CA, USA,
  2016. USENIX Association.

\bibitem{banga1999resource}
Gaurav Banga, Peter Druschel, and Jeffrey~C Mogul.
\newblock Resource containers: a new facility for resource management in server
  systems.
\newblock In {\em Proceedings of the third symposium on Operating systems
  design and implementation}, pages 45--58. USENIX Association, 1999.

\bibitem{barbalacetowards}
Antonio Barbalace, Alastair Murray, Rob Lyerly, and Binoy Ravindran.
\newblock Towards operating system support for heterogeneous-isa platforms.
\newblock In {\em Proceedings of The 4th Workshop on Systems for Future
  Multicore Architectures (4th SFMA)}, 2014.

\bibitem{Barbalace:2015:popcorn}
Antonio Barbalace, Marina Sadini, Saif Ansary, Christopher Jelesnianski, Akshay
  Ravichandran, Cagil Kendir, Alastair Murray, and Binoy Ravindran.
\newblock Popcorn: Bridging the programmability gap in heterogeneous-isa
  platforms.
\newblock In {\em Proceedings of the Tenth European Conference on Computer
  Systems}, EuroSys '15, pages 29:1--29:16, New York, NY, USA, 2015. ACM.

\bibitem{barham2003xen}
Paul Barham, Boris Dragovic, Keir Fraser, Steven Hand, Tim Harris, Alex Ho,
  Rolf Neugebauer, Ian Pratt, and Andrew Warfield.
\newblock Xen and the art of virtualization.
\newblock {\em ACM SIGOPS Operating Systems Review}, 37(5):164--177, 2003.

\bibitem{barroso2013datacenter}
Luiz~Andr{\'e} Barroso, Jimmy Clidaras, and Urs H{\"o}lzle.
\newblock The datacenter as a computer: an introduction to the design of
  warehouse-scale machines.
\newblock {\em Synthesis lectures on computer architecture}, 8(3):1--154, 2013.

\bibitem{barroso2009datacenter}
Luiz~Andr{\'e} Barroso and Urs H{\"o}lzle.
\newblock The datacenter as a computer: An introduction to the design of
  warehouse-scale machines.
\newblock {\em Synthesis lectures on computer architecture}, 4(1):1--108, 2009.

\bibitem{batlivala1992experience}
Nariman Batlivala, Barry Gleeson, James~R Hamrick, Scott Lurndal, Darren Price,
  James Soddy, and Vadim Abrossimov.
\newblock Experience with svr4 over chorus.
\newblock In {\em Proceedings of the Workshop on Micro-kernels and Other Kernel
  Architectures}, pages 223--242. USENIX Association, 1992.

\bibitem{Baumann:2009:multikernel}
Andrew Baumann, Paul Barham, Pierre-Evariste Dagand, Tim Harris, Rebecca
  Isaacs, Simon Peter, Timothy Roscoe, Adrian Sch{\"u}pbach, and Akhilesh
  Singhania.
\newblock The multikernel: a new os architecture for scalable multicore
  systems.
\newblock In {\em Proceedings of the ACM SIGOPS 22nd symposium on Operating
  systems principles}, pages 29--44. ACM, 2009.

\bibitem{beckmann2014pika}
Nathan~Z Beckmann, Charles Gruenwald~III, Christopher~R Johnson, Harshad
  Kasture, Filippo Sironi, Anant Agarwal, M~Frans Kaashoek, and Nickolai
  Zeldovich.
\newblock Pika: A network service for multikernel operating systems.
\newblock {\em Technical ReportMIT-CSAIL-TR-2014-002, MIT.}, 2014.

\bibitem{Belay:2012:DSU:2387880.2387913}
Adam Belay, Andrea Bittau, Ali Mashtizadeh, David Terei, David Mazi\`{e}res,
  and Christos Kozyrakis.
\newblock Dune: Safe user-level access to privileged cpu features.
\newblock In {\em Proceedings of the 10th USENIX Conference on Operating
  Systems Design and Implementation}, OSDI'12, pages 335--348. USENIX
  Association, 2012.

\bibitem{belay2014ix}
Adam Belay, George Prekas, Ana Klimovic, Samuel Grossman, Christos Kozyrakis,
  and Edouard Bugnion.
\newblock Ix: A protected dataplane operating system for high throughput and
  low latency.
\newblock In {\em 11th USENIX Symposium on Operating Systems Design and
  Implementation (OSDI 14),(Broomfield, CO)}, pages 49--65, 2014.

\bibitem{bienia2008parsec}
Christian Bienia, Sanjeev Kumar, Jaswinder~Pal Singh, and Kai Li.
\newblock The {PARSEC} benchmark suite: characterization and architectural
  implications.
\newblock In {\em Proceedings of the 17th international conference on Parallel
  architectures and compilation techniques}, pages 72--81. ACM, 2008.

\bibitem{boyd2008corey}
Silas Boyd-Wickizer, Haibo Chen, Rong Chen, Yandong Mao, Frans Kaashoek, Robert
  Morris, Aleksey Pesterev, Lex Stein, Ming Wu, Yuehua Dai, Yang Zhang, and
  Zheng Zhang.
\newblock Corey: An operating system for many cores.
\newblock In {\em Proceedings of the 8th USENIX Conference on Operating Systems
  Design and Implementation}, OSDI'08, pages 43--57, Berkeley, CA, USA, 2008.
  USENIX Association.

\bibitem{Boyd-Wickizer:2010:MOSBench}
Silas Boyd-Wickizer, Austin~T. Clements, Yandong Mao, Aleksey Pesterev,
  M.~Frans Kaashoek, Robert Morris, and Nickolai Zeldovich.
\newblock An analysis of linux scalability to many cores.
\newblock In {\em Proceedings of the 9th USENIX Conference on Operating Systems
  Design and Implementation}, OSDI'10, pages 1--8, Berkeley, CA, USA, 2010.
  USENIX Association.

\bibitem{bugnion1997disco}
Edouard Bugnion, Scott Devine, Kinshuk Govil, and Mendel Rosenblum.
\newblock Disco: Running commodity operating systems on scalable
  multiprocessors.
\newblock {\em ACM Transactions on Computer Systems (TOCS)}, 15(4):412--447,
  1997.

\bibitem{Chase:1994:SPS:195792.195795}
Jeffrey~S. Chase, Henry~M. Levy, Michael~J. Feeley, and Edward~D. Lazowska.
\newblock Sharing and protection in a single-address-space operating system.
\newblock {\em ACM Trans. Comput. Syst.}, 12(4):271--307, November 1994.

\bibitem{chen2013towards}
Chen Chen, Petros Maniatis, Adrian Perrig, Amit Vasudevan, and Vyas Sekar.
\newblock Towards verifiable resource accounting for outsourced computation.
\newblock In {\em ACM SIGPLAN Notices}, volume~48, pages 167--178. ACM, 2013.

\bibitem{chisnall2008definitive}
David Chisnall.
\newblock {\em The definitive guide to the xen hypervisor}.
\newblock Pearson Education, 2008.

\bibitem{clements2015scalable}
Austin~T Clements, M~Frans Kaashoek, Nickolai Zeldovich, Robert~T Morris, and
  Eddie Kohler.
\newblock The scalable commutativity rule: Designing scalable software for
  multicore processors.
\newblock {\em ACM Transactions on Computer Systems (TOCS)}, 32(4):10, 2015.

\bibitem{Colmenares:2013:TRO:2463209.2488827}
Juan~A. Colmenares, Gage Eads, Steven Hofmeyr, Sarah Bird, Miquel Moret\'{o},
  David Chou, Brian Gluzman, Eric Roman, Davide~B. Bartolini, Nitesh Mor, Krste
  Asanovi\'{c}, and John~D. Kubiatowicz.
\newblock Tessellation: Refactoring the os around explicit resource containers
  with continuous adaptation.
\newblock In {\em Proceedings of the 50th Annual Design Automation Conference},
  DAC '13, pages 76:1--76:10, New York, NY, USA, 2013. ACM.

\bibitem{dean2013tail}
Jeffrey Dean and Luiz~Andr{\'e} Barroso.
\newblock The tail at scale.
\newblock {\em Communications of the ACM}, 56(2):74--80, 2013.

\bibitem{Deng:2017:DWT}
Liang Deng, Peng Liu, Jun Xu, Ping Chen, and Qingkai Zeng.
\newblock Dancing with wolves: Towards practical event-driven vmm monitoring.
\newblock In {\em Proceedings of the 13th ACM SIGPLAN/SIGOPS International
  Conference on Virtual Execution Environments}, VEE '17, pages 83--96, New
  York, NY, USA, 2017. ACM.

\bibitem{Elphinstone:2013:LSW:2517349.2522720}
Kevin Elphinstone and Gernot Heiser.
\newblock From l3 to sel4 what have we learnt in 20 years of l4 microkernels?
\newblock In {\em Proceedings of the Twenty-Fourth ACM Symposium on Operating
  Systems Principles}, SOSP '13, pages 133--150, New York, NY, USA, 2013. ACM.

\bibitem{Engler:1995:EOS:224056.224076}
D.~R. Engler, M.~F. Kaashoek, and J.~O'Toole, Jr.
\newblock Exokernel: An operating system architecture for application-level
  resource management.
\newblock In {\em Proceedings of the Fifteenth ACM Symposium on Operating
  Systems Principles}, SOSP '95, pages 251--266, New York, NY, USA, 1995. ACM.

\bibitem{engler1995exokernel}
Dawson~R Engler, M~Frans Kaashoek, et~al.
\newblock {\em Exokernel: An operating system architecture for
  application-level resource management}, volume~29.
\newblock ACM, 1995.

\bibitem{Ford:1996:MMR:238721.238769}
Bryan Ford, Mike Hibler, Jay Lepreau, Patrick Tullmann, Godmar Back, and
  Stephen Clawson.
\newblock Microkernels meet recursive virtual machines.
\newblock In {\em Proceedings of the Second USENIX Symposium on Operating
  Systems Design and Implementation}, OSDI '96, pages 137--151. ACM, 1996.

\bibitem{gabber1999pebble}
Eran Gabber, Christopher Small, John~L Bruno, Jos{\'e}~Carlos Brustoloni, and
  Abraham Silberschatz.
\newblock The pebble component-based operating system.
\newblock In {\em USENIX Annual Technical Conference, General Track}, pages
  267--282, 1999.

\bibitem{gamsa1999tornado}
Ben Gamsa, Orran Krieger, Jonathan Appavoo, and Michael Stumm.
\newblock Tornado: Maximizing locality and concurrency in a shared memory
  multiprocessor operating system.
\newblock In {\em Proceedings of the Third Symposium on Operating Systems
  Design and Implementation}, OSDI '99, pages 87--100, Berkeley, CA, USA, 1999.
  USENIX Association.

\bibitem{gefflaut2000sawmill}
Alain Gefflaut, Trent Jaeger, Yoonho Park, Jochen Liedtke, Kevin~J Elphinstone,
  Volkmar Uhlig, Jonathon~E Tidswell, Luke Deller, and Lars Reuther.
\newblock The sawmill multiserver approach.
\newblock In {\em Proceedings of the 9th workshop on ACM SIGOPS European
  workshop: beyond the PC: new challenges for the operating system}, pages
  109--114. ACM, 2000.

\bibitem{Ghodsi:2011:DRF:1972457.1972490}
Ali Ghodsi, Matei Zaharia, Benjamin Hindman, Andy Konwinski, Scott Shenker, and
  Ion Stoica.
\newblock Dominant resource fairness: Fair allocation of multiple resource
  types.
\newblock In {\em Proceedings of the 8th USENIX Conference on Networked Systems
  Design and Implementation}, NSDI'11, pages 323--336, Berkeley, CA, USA, 2011.
  USENIX Association.

\bibitem{goldberg1974survey}
Robert~P Goldberg.
\newblock Survey of virtual machine research.
\newblock {\em Computer}, 7(6):34--45, 1974.

\bibitem{golub1990unix}
David~B Golub, Randall~W Dean, Alessandro Forin, and Richard~F Rashid.
\newblock Unix as an application program.
\newblock In {\em UsENIX summer}, pages 87--95, 1990.

\bibitem{gough2007kernel}
Corey Gough, Suresh Siddha, and Ken Chen.
\newblock Kernel scalability---expanding the horizon beyond fine grain locks.
\newblock In {\em Proceedings of the Linux Symposium}, pages 153--165, 2007.

\bibitem{govil1999cellular}
Kinshuk Govil, Dan Teodosiu, Yongqiang Huang, and Mendel Rosenblum.
\newblock Cellular disco: Resource management using virtual clusters on
  shared-memory multiprocessors.
\newblock In {\em ACM SIGOPS Operating Systems Review}, volume~33, pages
  154--169. ACM, 1999.

\bibitem{Graham:1968:PIP:363095.363146}
Robert~M. Graham.
\newblock Protection in an information processing utility.
\newblock {\em Commun. ACM}, 11(5):365--369, May 1968.

\bibitem{TPM}
Trusted~Computing Group.
\newblock Trusted platform module. {Accessed April 2017.}
\newblock \url{http://www.trustedcomputinggroup.org/}.

\bibitem{Gulati:2010:MHT:1924943.1924974}
Ajay Gulati, Arif Merchant, and Peter~J. Varman.
\newblock mclock: Handling throughput variability for hypervisor io scheduling.
\newblock In {\em Proceedings of the 9th USENIX Conference on Operating Systems
  Design and Implementation}, OSDI'10, pages 1--7, Berkeley, CA, USA, 2010.
  USENIX Association.

\bibitem{guniguntala2008read}
Dinakar Guniguntala, Paul~E McKenney, Josh Triplett, and Jonathan Walpole.
\newblock The read-copy-update mechanism for supporting real-time applications
  on shared-memory multiprocessor systems with linux.
\newblock {\em IBM Systems Journal}, 47(2):221--236, 2008.

\bibitem{henning:2006:speccpu}
John~L Henning.
\newblock Spec cpu2006 benchmark descriptions.
\newblock {\em ACM SIGARCH Computer Architecture News}, 34(4):1--17, 2006.

\bibitem{herder2006minix}
Jorrit~N Herder, Herbert Bos, Ben Gras, Philip Homburg, and Andrew~S Tanenbaum.
\newblock Minix 3: A highly reliable, self-repairing operating system.
\newblock {\em ACM SIGOPS Operating Systems Review}, 40(3):80--89, 2006.

\bibitem{Howry:1972:MSP:850614.850619}
Sam Howry.
\newblock A multiprogramming system for process control.
\newblock {\em SIGOPS Oper. Syst. Rev.}, 6(1/2):24--30, June 1972.

\bibitem{Huang:2016:ESL}
Jian Huang, Moinuddin~K. Qureshi, and Karsten Schwan.
\newblock An evolutionary study of linux memory management for fun and profit.
\newblock In {\em Proceedings of the 2016 USENIX Conference on Usenix Annual
  Technical Conference}, USENIX ATC '16, pages 465--478, Berkeley, CA, USA,
  2016. USENIX Association.

\bibitem{jalaparti2013speeding}
Virajith Jalaparti, Peter Bodik, Srikanth Kandula, Ishai Menache, Mikhail
  Rybalkin, and Chenyu Yan.
\newblock Speeding up distributed request-response workflows.
\newblock In {\em Proceedings of the ACM SIGCOMM 2013 conference on SIGCOMM},
  pages 219--230. ACM, 2013.

\bibitem{janapa2010web}
Vijay Janapa~Reddi, Benjamin~C Lee, Trishul Chilimbi, and Kushagra Vaid.
\newblock Web search using mobile cores: quantifying and mitigating the price
  of efficiency.
\newblock {\em ACM SIGARCH Computer Architecture News}, 38(3):314--325, 2010.

\bibitem{Jann:2003:IBM_LPARs}
J.~Jann, L.~M. Browning, and R.~S. Burugula.
\newblock Dynamic reconfiguration: Basic building blocks for autonomic
  computing on ibm pseries servers.
\newblock {\em IBM Syst. J.}, 42(1):29--37, January 2003.

\bibitem{jones:1996:netperf}
Rick Jones et~al.
\newblock Netperf: a network performance benchmark.
\newblock {\em Information Networks Division, Hewlett-Packard Company}, 1996.

\bibitem{julin1995mach}
J~Mark Stevenson Daniel~P Julin.
\newblock Mach-us: Unix on generic os object servers.
\newblock In {\em Proceedings of the... USENIX Technical Conference}, page 119.
  Usenix Assoc, 1995.

\bibitem{kale2011distributing}
Amit Kale, Parag Mittal, Shekhar Manek, Neha Gundecha, and Madhuri Londhe.
\newblock Distributing subsystems across different kernels running
  simultaneously in a multi-core architecture.
\newblock In {\em Computational Science and Engineering (CSE), 2011 IEEE 14th
  International Conference on}, pages 114--120. IEEE, 2011.

\bibitem{kanev2015profiling}
Svilen Kanev, Juan~Pablo Darago, Kim Hazelwood, Parthasarathy Ranganathan, Tipp
  Moseley, Gu-Yeon Wei, and David Brooks.
\newblock Profiling a warehouse-scale computer.
\newblock In {\em Computer Architecture (ISCA), 2015 ACM/IEEE 42nd Annual
  International Symposium on}, pages 158--169. IEEE, 2015.

\bibitem{kasture2014ubik}
Harshad Kasture and Daniel Sanchez.
\newblock Ubik: efficient cache sharing with strict qos for latency-critical
  workloads.
\newblock In {\em Proceedings of the 19th international conference on
  Architectural support for programming languages and operating systems}, pages
  729--742. ACM, 2014.

\bibitem{keller2010nohype}
Eric Keller, Jakub Szefer, Jennifer Rexford, and Ruby~B Lee.
\newblock Nohype: virtualized cloud infrastructure without the virtualization.
\newblock In {\em ACM SIGARCH Computer Architecture News}, volume~38, pages
  350--361. ACM, 2010.

\bibitem{kivity2007kvm}
Avi Kivity, Yaniv Kamay, Dor Laor, Uri Lublin, and Anthony Liguori.
\newblock kvm: the linux virtual machine monitor.
\newblock In {\em Proceedings of the Linux Symposium}, volume~1, pages
  225--230, 2007.

\bibitem{kooburat2011best}
Thawan Kooburat and Michael Swift.
\newblock The best of both worlds with on-demand virtualization.
\newblock In {\em Workshop on Hot Topics in Operating Systems (HotOS)}, 2011.

\bibitem{tun_site}
Maxixm Krasnyansky and Yevmenkin Maksim.
\newblock Universal {TUN/TAP} driver. {Accessed Dec 2014.}
\newblock \url{http://vtun.sourceforge.net/tun}.

\bibitem{krieger2006k42}
Orran Krieger, Marc Auslander, Bryan Rosenburg, Robert~W Wisniewski, Jimi
  Xenidis, Dilma Da~Silva, Michal Ostrowski, Jonathan Appavoo, Maria Butrico,
  Mark Mergen, et~al.
\newblock K42: building a complete operating system.
\newblock {\em ACM SIGOPS Operating Systems Review}, 40(4):133--145, 2006.

\bibitem{kroah2001deivcehotplug}
Greg Kroah-Hartman.
\newblock Hotpluggable devices and the linux kernel.
\newblock Ottawa Linux Symposium, 2001.

\bibitem{leverich2014reconciling}
Jacob Leverich and Christos Kozyrakis.
\newblock Reconciling high server utilization and sub-millisecond
  quality-of-service.
\newblock In {\em Proceedings of the Ninth European Conference on Computer
  Systems}, page~4. ACM, 2014.

\bibitem{litales}
Jialin Li, Naveen~Kr Sharma, Dan~RK Ports, and Steven~D Gribble.
\newblock Tales of the tail: Hardware, os, and application-level sources of
  tail latency.
\newblock In {\em Proceedings of the ACM Symposium on Cloud Computing}, pages
  1--14. ACM, 2014.

\bibitem{lin2014k2}
Felix~Xiaozhu Lin, Zhen Wang, and Lin Zhong.
\newblock K2: A mobile operating system for heterogeneous coherence domains.
\newblock In {\em Proceedings of the 19th international conference on
  Architectural support for programming languages and operating systems}, pages
  285--300. ACM, 2014.

\bibitem{liu2006high}
Jiuxing Liu, Wei Huang, B{\"u}lent Abali, and Dhabaleswar~K Panda.
\newblock High performance vmm-bypass i/o in virtual machines.
\newblock In {\em USENIX Annual Technical Conference, General Track}, pages
  29--42, 2006.

\bibitem{Madhavapeddy:2013:ULO:2451116.2451167}
Anil Madhavapeddy, Richard Mortier, Charalampos Rotsos, David Scott, Balraj
  Singh, Thomas Gazagnaire, Steven Smith, Steven Hand, and Jon Crowcroft.
\newblock Unikernels: Library operating systems for the cloud.
\newblock In {\em Proceedings of the Eighteenth International Conference on
  Architectural Support for Programming Languages and Operating Systems},
  ASPLOS '13, pages 461--472, New York, NY, USA, 2013. ACM.

\bibitem{mellor1991algorithms}
John~M Mellor-Crummey and Michael~L Scott.
\newblock Algorithms for scalable synchronization on shared-memory
  multiprocessors.
\newblock {\em ACM Transactions on Computer Systems (TOCS)}, 9(1):21--65, 1991.

\bibitem{merkel2014docker}
Dirk Merkel.
\newblock Docker: lightweight linux containers for consistent development and
  deployment.
\newblock {\em Linux Journal}, 2014(239):2, 2014.

\bibitem{Michael:2004:SLD}
Maged~M. Michael.
\newblock Scalable lock-free dynamic memory allocation.
\newblock In {\em Proceedings of the ACM SIGPLAN 2004 Conference on Programming
  Language Design and Implementation}, PLDI '04, pages 35--46, New York, NY,
  USA, 2004. ACM.

\bibitem{mogul2013nic}
Jeffrey~C Mogul, Jayaram Mudigonda, Jose~Renato Santos, and Yoshio Turner.
\newblock The nic is the hypervisor: bare-metal guests in iaas clouds.
\newblock In {\em Proceedings of the 14th USENIX conference on Hot Topics in
  Operating Systems}, pages 2--2. USENIX Association, 2013.

\bibitem{Mucci:1998:cachebench}
Philip Mucci and Kevin~S. London.
\newblock Low level architectural characterization benchmarks for parallel
  computers.
\newblock Technical report, 1998.

\bibitem{mwaikambo2004linuxCPUhotplug}
Zwane Mwaikambo, Ashok Raj, Rusty Russell, Joel Schopp, and Srivatsa Vaddagiri.
\newblock Linux kernel hotplug cpu support.
\newblock In {\em Linux Symposium}, volume~2, 2004.

\bibitem{nightingale2009helios}
Edmund~B Nightingale, Orion Hodson, Ross McIlroy, Chris Hawblitzel, and Galen
  Hunt.
\newblock Helios: heterogeneous multiprocessing with satellite kernels.
\newblock In {\em Proceedings of the ACM SIGOPS 22nd symposium on Operating
  systems principles}, pages 221--234. ACM, 2009.

\bibitem{nikolaev2013virtuos}
Ruslan Nikolaev and Godmar Back.
\newblock Virtuos: an operating system with kernel virtualization.
\newblock In {\em Proceedings of the Twenty-Fourth ACM Symposium on Operating
  Systems Principles}, pages 116--132. ACM, 2013.

\bibitem{Nitu:2017:SBQ}
Vlad Nitu, Pierre Olivier, Alain Tchana, Daniel Chiba, Antonio Barbalace,
  Daniel Hagimont, and Binoy Ravindran.
\newblock Swift birth and quick death: Enabling fast parallel guest boot and
  destruction in the xen hypervisor.
\newblock In {\em Proceedings of the 13th ACM SIGPLAN/SIGOPS International
  Conference on Virtual Execution Environments}, VEE '17, pages 1--14, New
  York, NY, USA, 2017. ACM.

\bibitem{nomura2011mint}
Yoshinari Nomura, Ryota Senzaki, Daiki Nakahara, Hiroshi Ushio, Tetsuya
  Kataoka, and Hideo Taniguchi.
\newblock Mint: Booting multiple linux kernels on a multicore processor.
\newblock In {\em Broadband and Wireless Computing, Communication and
  Applications (BWCCA), 2011 International Conference on}, pages 555--560.
  IEEE, 2011.

\bibitem{pavlo2009comparison}
Andrew Pavlo, Erik Paulson, Alexander Rasin, Daniel~J Abadi, David~J DeWitt,
  Samuel Madden, and Michael Stonebraker.
\newblock A comparison of approaches to large-scale data analysis.
\newblock In {\em Proceedings of the 2009 ACM SIGMOD International Conference
  on Management of data}, pages 165--178. ACM, 2009.

\bibitem{peter2014arrakis}
Simon Peter, Jialin Li, Irene Zhang, Dan~RK Ports, Doug Woos, Arvind
  Krishnamurthy, Thomas Anderson, and Timothy Roscoe.
\newblock Arrakis: the operating system is the control plane.
\newblock In {\em Proceedings of the 11th USENIX conference on Operating
  Systems Design and Implementation}, pages 1--16. USENIX Association, 2014.

\bibitem{ravindranath2013timecard}
Lenin Ravindranath, Jitendra Padhye, Ratul Mahajan, and Hari Balakrishnan.
\newblock Timecard: controlling user-perceived delays in server-based mobile
  applications.
\newblock In {\em Proceedings of the Twenty-Fourth ACM Symposium on Operating
  Systems Principles}, pages 85--100. ACM, 2013.

\bibitem{russinovich2008inside}
Mark Russinovich.
\newblock Inside windows server 2008 kernel changes.
\newblock {\em Microsoft TechNet Magazine}, 2008.

\bibitem{Saltzer:1974:PCI:361011.361067}
Jerome~H. Saltzer.
\newblock Protection and the control of information sharing in multics.
\newblock {\em Commun. ACM}, 17(7):388--402, July 1974.

\bibitem{schopp2006memoryhotplug}
Joel~H Schopp, Keir Fraser, and Martine~J Silbermann.
\newblock Resizing memory with balloons and hotplug.
\newblock In {\em Proceedings of the Linux Symposium}, volume~2, pages
  313--319, 2006.

\bibitem{schupbach2008embracing}
Adrian Sch{\"u}pbach, Simon Peter, Andrew Baumann, Timothy Roscoe, Paul Barham,
  Tim Harris, and Rebecca Isaacs.
\newblock Embracing diversity in the barrelfish manycore operating system.
\newblock In {\em Proceedings of the Workshop on Managed Many-Core Systems},
  page~27, 2008.

\bibitem{sekar2011verifiable}
Vyas Sekar and Petros Maniatis.
\newblock Verifiable resource accounting for cloud computing services.
\newblock In {\em Proceedings of the 3rd ACM workshop on Cloud computing
  security workshop}, pages 21--26. ACM, 2011.

\bibitem{shimosawa2008logical}
Taku Shimosawa, Hiroya Matsuba, and Yutaka Ishikawa.
\newblock Logical partitioning without architectural supports.
\newblock In {\em Computer Software and Applications, 2008. COMPSAC'08. 32nd
  Annual IEEE International}, pages 355--364. IEEE, 2008.

\bibitem{shue2012performance}
David Shue, Michael~J Freedman, and Anees Shaikh.
\newblock Performance isolation and fairness for multi-tenant cloud storage.
\newblock In {\em OSDI}, volume~12, pages 349--362, 2012.

\bibitem{soltesz2007container}
Stephen Soltesz, Herbert P{\"o}tzl, Marc~E Fiuczynski, Andy Bavier, and Larry
  Peterson.
\newblock Container-based operating system virtualization: a scalable,
  high-performance alternative to hypervisors.
\newblock In {\em ACM SIGOPS Operating Systems Review}, volume~41, pages
  275--287. ACM, 2007.

\bibitem{staelin2002lmbench3}
Carl Staelin.
\newblock lmbench3: measuring scalability.
\newblock Technical report, Citeseer, 2002.

\bibitem{szefer2011eliminating}
Jakub Szefer, Eric Keller, Ruby~B Lee, and Jennifer Rexford.
\newblock Eliminating the hypervisor attack surface for a more secure cloud.
\newblock In {\em Proceedings of the 18th ACM conference on Computer and
  communications security}, pages 401--412. ACM, 2011.

\bibitem{fuse_site}
Miklos Szeredi.
\newblock {FUSE}: {F}ilesystem in {U}serspace. {Accessed Dec 2014.}
\newblock \url{http://fuse.sourceforge.net/}.

\bibitem{Tang:2011:IMS}
Lingjia Tang, Jason Mars, Neil Vachharajani, Robert Hundt, and Mary~Lou Soffa.
\newblock The impact of memory subsystem resource sharing on datacenter
  applications.
\newblock In {\em Proceedings of the 38th Annual International Symposium on
  Computer Architecture}, ISCA '11, pages 283--294, New York, NY, USA, 2011.
  ACM.

\bibitem{Tsai:2015:GMV}
Chia-Che Tsai, Yang Zhan, Jayashree Reddy, Yizheng Jiao, Tao Zhang, and
  Donald~E. Porter.
\newblock How to get more value from your file system directory cache.
\newblock In {\em Proceedings of the 25th Symposium on Operating Systems
  Principles}, SOSP '15, pages 441--456, New York, NY, USA, 2015. ACM.

\bibitem{unrau1995hierarchical}
Ronald~C Unrau, Orran Krieger, Benjamin Gamsa, and Michael Stumm.
\newblock Hierarchical clustering: A structure for scalable multiprocessor
  operating system design.
\newblock {\em The Journal of Supercomputing}, 9(1-2):105--134, 1995.

\bibitem{verghese1998performance}
Ben Verghese, Anoop Gupta, and Mendel Rosenblum.
\newblock Performance isolation: sharing and isolation in shared-memory
  multiprocessors.
\newblock In {\em ACM SIGPLAN Notices}, volume~33, pages 181--192. ACM, 1998.

\bibitem{DiDi2011}
C.~Villavieja, V.~Karakostas, L.~Vilanova, Y.~Etsion, A.~Ramirez, A.~Mendelson,
  N.~Navarro, A.~Cristal, and O.~S. Unsal.
\newblock Didi: Mitigating the performance impact of tlb shootdowns using a
  shared tlb directory.
\newblock In {\em 2011 International Conference on Parallel Architectures and
  Compilation Techniques}, pages 340--349, Oct 2011.

\bibitem{vulimiri2012more}
Ashish Vulimiri, Oliver Michel, P~Godfrey, and Scott Shenker.
\newblock More is less: reducing latency via redundancy.
\newblock In {\em Proceedings of the 11th ACM Workshop on Hot Topics in
  Networks}, pages 13--18. ACM, 2012.

\bibitem{Wang:2014:BigDataBench}
Lei Wang, Jianfeng Zhan, Chunjie Luo, Yuqing Zhu, Qiang Yang, Yongqiang He,
  Wanling Gao, Zhen Jia, Yingjie Shi, Shujie Zhang, Chen Zheng, Gang Lu,
  K.~Zhan, Xiaona Li, and Bizhu Qiu.
\newblock Bigdatabench: A big data benchmark suite from internet services.
\newblock In {\em High Performance Computer Architecture (HPCA), 2014 IEEE 20th
  International Symposium on}, pages 488--499, Feb 2014.

\bibitem{wentzlaff2009factored}
David Wentzlaff and Anant Agarwal.
\newblock Factored operating systems (fos): the case for a scalable operating
  system for multicores.
\newblock {\em ACM SIGOPS Operating Systems Review}, 43(2):76--85, 2009.

\bibitem{zaharia2012resilient}
Matei Zaharia, Mosharaf Chowdhury, Tathagata Das, Ankur Dave, Justin Ma, Murphy
  McCauley, Michael~J Franklin, Scott Shenker, and Ion Stoica.
\newblock Resilient distributed datasets: A fault-tolerant abstraction for
  in-memory cluster computing.
\newblock In {\em Proceedings of the 9th USENIX conference on Networked Systems
  Design and Implementation}, pages 2--2. USENIX Association, 2012.

\bibitem{Zellweger:2014:DecouplingCores}
Gerd Zellweger, Simon Gerber, Kornilios Kourtis, and Timothy Roscoe.
\newblock Decoupling cores, kernels, and operating systems.
\newblock In {\em Proceedings of the 11th USENIX Conference on Operating
  Systems Design and Implementation}, OSDI'14, pages 17--31, Berkeley, CA, USA,
  2014. USENIX Association.

\bibitem{zhang2011cloudvisor}
Fengzhe Zhang, Jin Chen, Haibo Chen, and Binyu Zang.
\newblock Cloudvisor: retrofitting protection of virtual machines in
  multi-tenant cloud with nested virtualization.
\newblock In {\em Proceedings of the Twenty-Third ACM Symposium on Operating
  Systems Principles}, pages 203--216. ACM, 2011.

\bibitem{zhang2013cpi}
Xiao Zhang, Eric Tune, Robert Hagmann, Rohit Jnagal, Vrigo Gokhale, and John
  Wilkes.
\newblock Cpi 2: Cpu performance isolation for shared compute clusters.
\newblock In {\em Proceedings of the 8th ACM European Conference on Computer
  Systems}, pages 379--391. ACM, 2013.

\bibitem{Zhong:2012:OXH}
Alin Zhong, Hai Jin, Song Wu, Xuanhua Shi, and Wei Gen.
\newblock Optimizing xen hypervisor by using lock-aware scheduling.
\newblock In {\em Proceedings of the 2012 Second International Conference on
  Cloud and Green Computing}, CGC '12, pages 31--38, Washington, DC, USA, 2012.
  IEEE Computer Society.

\end{thebibliography}

\end{document}